\documentclass[10pt, conference]{IEEEtran}

\usepackage{adjustbox}
\usepackage{subcaption}
\usepackage{caption}
\usepackage{graphicx} 
\usepackage{stfloats}
\usepackage{booktabs}
\usepackage{multirow}
\usepackage{optidef}

\usepackage{tabularx}
\usepackage{booktabs}

\IEEEoverridecommandlockouts
\usepackage[percent]{overpic}
\usepackage{graphicx}
\usepackage{amsmath,amssymb,amsfonts}
\usepackage{float}
\usepackage{multicol}
\usepackage{lipsum} 
\usepackage{textcomp}
\usepackage{xcolor}
\usepackage{breakurl}
\usepackage{xurl}
\usepackage{longtable} 
\usepackage{array}     
\usepackage{verbatim}
\usepackage[ruled, vlined, linesnumbered, boxed]{algorithm2e}
\usepackage{algpseudocode}

\title{{\em ProFlow}: RL-Driven and Performance-Aware Proactive Flow Placement in Datacenter Networks\thanks{This material is based upon work supported by the National Science Foundation (NSF) under Award Number CNS-2232889.}}

\author{
Sourya Saha, Md Nurul Absur, Saptarshi Debroy\\
City University of New York\\ 
Emails: \textit{ssaha2@gradcenter.cuny.edu, mabsur@gradcenter.cuny.edu, saptarshi.debroy@hunter.cuny.edu}
}

\begin{document}

\maketitle
\thispagestyle{empty}
\pagestyle{empty}
\maketitle

\begin{abstract}
In datacenter fabrics composed of leaf and aggregation switches, competing flows may become co-located on shared aggregation switches, creating congestion that can significantly degrade protected flows. However, before throughput degradation becomes observable, the network often exhibits early signs characterized by rising flow activity and queue overflow signals. Existing congestion-management approaches primarily react only after congestion becomes visible, leaving these early signs largely unexploited. In this paper, we propose \emph{ProFlow}, a proactive flow-placement framework for protecting performance-sensitive traffic in multi-tenant datacenter networks, thereby utilizing the early signs of potential throughput degradations. \emph{ProFlow} leverages distributed telemetry signals and offline-trained reinforcement learning (RL) to identify precursor congestion conditions and proactively reroute protected flows before throughput degradation occurs. Evaluation results using FABRIC testbed show that \emph{ProFlow} achieves approximately 40\% higher mean throughput than a reactive rerouting baseline while initiating rerouting decisions around 34 seconds earlier on average, demonstrating the effectiveness of anticipatory congestion management. 
\end{abstract}

\begin{IEEEkeywords}
Datacenter networks, proactive flow placement, congestion management, reinforcement learning, software-defined networking, multi-tenant networks
\end{IEEEkeywords}


\section{Introduction}
\label{sec:intro}

Modern cloud datacenters host a diverse mix of workloads, including distributed training jobs, large-scale data processing pipelines, and latency-sensitive services, all placing significant network demand~\cite{multi-cloud,adon,icnp}. Many of these applications depend on sustained communication efficiency, and even short-lived throughput degradation can slow end-to-end execution by delaying synchronization and data exchange \cite{EfficientSparseCollectiveCommunication,veca}. To support such workloads at scale, datacenter operators widely rely on leaf-aggregate network fabrics \cite{AlFares}, in which servers connect to leaf switches that forward traffic through a set of shared aggregation switches, providing multiple parallel paths between endpoints. While this architecture improves path diversity and network capacity, it also creates a fundamental challenge in multi-tenant environments, as in, flows from different tenants may be placed on the same aggregation switch and therefore contend for shared bandwidth. In practice, common path selection mechanisms do not reason about the future congestion impact of individual flow placements, which can allow harmful traffic co-location to persist. As a result, tenants with stricter service-level objectives may experience disproportionate performance degradation when their flows are co-located with aggressive or bandwidth-intensive traffic~\cite{vectrust}. Ensuring that such flows maintain stable throughput becomes critical for both application correctness and SLA compliance, making intelligent flow placement across shared aggregation switches a key requirement in modern datacenters~\cite{measurement,pca}.

\begin{figure}[tb]
  \centering
  \includegraphics[width=0.90\columnwidth]{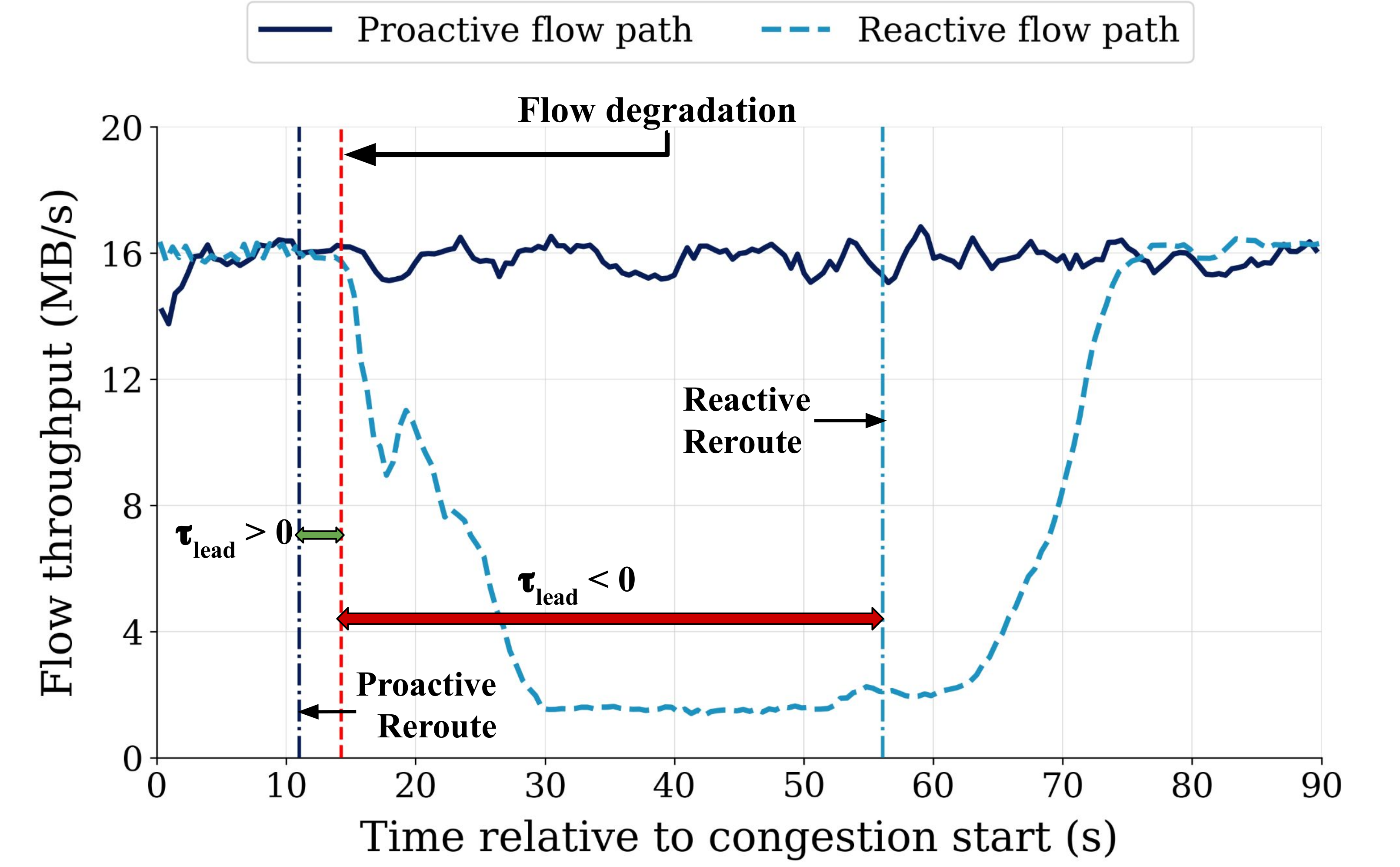}
  \caption{Temporal dynamics of congestion onset and protected flow degradation in a multi-tenant leaf-aggregate fabric. Flow throughput (smoothed) shown over a 90-second window. The left axis marks the point at which congestion flows are launched; the {\color{red}red} dashed line marks the onset of protected flow degradation. A temporal gap of approximately 15 sec. separates these two events of flow degradation and congestion flow start; during this, congestion is already observable building, yet the protected flow remains healthy. The {\color{cyan}cyan} dotted line marks the point at which a reactive approach decides to reroute the flow by which time, the flow throughput has completely degraded.}
  \label{fig:temporal_gap}
  \vspace{-0.3in}
\end{figure}

While interference originates from the co-location of competing flows on shared aggregation switches, the challenge lies in identifying when such placement begins to pose a risk. In practice, bandwidth-intensive flows assigned to the same aggregation switch as a latency-sensitive protected flow can rapidly consume available capacity, eventually causing queue overflow and throughput degradation. However, as illustrated in Figure \ref{fig:temporal_gap}, manifestation of congestion is preceded by a precursor phase during which the network already exhibits early warning signals, such as increasing active flow counts and rising queue overflow activity, while the protected flow itself continues to operate normally. This creates a temporal gap between the emergence of congestion-inducing conditions and their observable impact on performance. By the time conventional indicators such as throughput collapse become visible, degradation is already underway, leaving limited opportunity for corrective action. {\em Exploiting this gap requires reasoning about whether current network conditions are likely to evolve into future congestion on a particular aggregation switch before any direct performance degradation is observed. Such reasoning is fundamentally challenging because the precursor signals are distributed, noisy, and highly dependent on the evolving global traffic state.}

Existing approaches to congestion management in datacenter networks do not fully address the problem of proactive flow placement under emerging congestion. A large class of transport-layer mechanisms improves network utilization and fairness by reacting to congestion signals using richer telemetry, faster control loops, or host-level feedback \cite{powertcp,bolt,hostcc}, but these approaches regulate transmission only after contention begins to manifest, without influencing where flows are placed in the fabric. At the network level, load balancing and repathing techniques redistribute traffic across available paths based on observed congestion \cite{plb}, yet their decisions are still triggered by conditions on the current path, rather than by anticipating which alternative paths may become congested. Architectural proposals aim to avoid congestion altogether through stronger network guarantees and redesigned fabric support \cite{harmony}, but such approaches require changes to the underlying infrastructure and are not easily deployable in existing systems. More recently, learning-based methods have explored reinforcement learning for datacenter control \cite{acc,auto}, demonstrating the ability to optimize congestion-control parameters or global traffic patterns, but not addressing proactive per-flow protection based on early precursor signals. {\em As a result, the problem of deciding whether a currently healthy flow should be moved before degradation begins, based on indirect and distributed signals of future congestion, remains largely unaddressed.}

In this paper, we present \emph{ProFlow}, a reinforcement learning (RL)-driven proactive flow-placement framework for multi-tenant datacenter networks. \emph{ProFlow} identifies when a well-performing flow is at risk of future degradation and proactively relocates it before congestion impacts throughput. It leverages distributed telemetry from leaf and aggregation switches to learn precursor patterns of future congestion, evaluates alternative aggregation paths, and selects the placement expected to maintain stable performance. The approach operates over existing network infrastructures without requiring changes to the underlying fabric, making it readily applicable to practical multi-tenant deployments.

We implement \emph{ProFlow} on a programmable leaf-aggregation datacenter environment deployed across multiple FABRIC testbed sites \cite{fabric-2019}. We evaluate it under realistic multi-tenant traffic scenarios in which protected flows coexist with competing bandwidth-intensive transfers that create dynamic and heterogeneous congestion across aggregation paths. We compare \emph{ProFlow} against static placement and reactive rerouting driven by observed congestion. Across these scenarios, \emph{ProFlow} preserves protected-flow throughput under contention, achieving approximately 40\% higher mean throughput than the reactive baseline while initiating reroutes about 34 seconds earlier on average. These results show that early precursor signals can support effective proactive placement before congestion becomes directly observable through throughput degradation.

The rest of the paper is organized as follows. Section \ref{sec:related} reviews related work. Section~ \ref{sec:background} provides an overview of the system design on which our work is based and the formulation of the problem we are dealing with in this work. Section \ref{sec:design} presents the design of \emph{ProFlow}. Section \ref{sec:eval-results} presents the system implementation and evaluates ProFlow under realistic multi-tenant traffic conditions. Finally, Section \ref{sec:conclusion} concludes the paper and outlines directions for future work.


\vspace{-0.2cm}

\section{Related Work}
\label{sec:related}

The current literature has highlighted congestion across different layers
\cite{10891783, 9370092, https://doi.org/10.1002/dac.70103}.\\ 
\noindent\textbf{Datacenter Congestion Control:}
Datacenter transport protocols have addressed the problem of maintaining high throughput \& low latency under heterogeneous \& bursty workloads. Early contributions, such as DCTCP, use 
ECN feedback to estimate congestion levels and adjust the congestion window accordingly \cite{10.1145/1851275.1851192}. This enables high throughput while retaining switch queues less crowded. More recent contributions address improving the speed and precision of congestion feedback. HPCC leverages in-network telemetry to acquire fine-grained link-road information and precisely control sending rates \cite{10.1145/3341302.3342085}.  PowerTCP \cite{10682942} adapts the congestion window using the bandwidth-window product feedback for rapid reaction under dynamic traffic conditions. These transport-layer mechanisms improve congestion response by observing ECN mark reflections, telemetry, delay, or queueing signals. Nevertheless, these works do not decide where the flow should be placed in the network fabric. To address this limitation, they regulate the sender once congestion pressure builds, but do not prevent a vulnerable flow from being co-located with congestion-inducing traffic on a shared aggregation path.\\ 
\noindent\textbf{Load Balancing and Congestion-Aware Path Selection:}
Distributing traffic across multipath paths in Clos \& leaf-aggregate datacenter fabrics is becoming more common in recent times. ECMP is widely used because of its simplicity and scalability, but its hash-based placement can cause runtime congestion and map competing flows onto the same bottleneck path \cite{fathurrohim2025evaluation}. Congestion-aware load balancing systems address the aforementioned limitations by using network feedback to shift traffic away from congested paths. Liu et al. explain that the advanced algorithm \emph{Gemma} leverages queue length to improve network performance in Remote Direct Memory Access (RDMA) \cite{11083214}. Some other recent works, such as CONGA \cite{das2025survey} and PLB \cite{10.1145/3544216.3544226}, propose path changes to reduce packet reordering. These systems demonstrate that path changes mitigate hotspots in multipath datacenter fabrics. However, their triggering status condition remains fundamentally reactive and susceptible to predictable activity. This makes the problem counterfactual rather than purely congestion-responsive.\\ 
\noindent\textbf{Learning-Based Datacenter Traffic Optimization:}
The emergence of machine learning has also been explored for data center traffic control. Reinforcement learning mitigates challenges such as partial observability, nonstationarity, and multi-objectivity \cite{10.1145/3512798.3512815}. Other learning-based control methods fine-tune congestion control parameters or optimize global traffic patterns \cite{pandove2024optimizing,boussaoud2025adaptive}. These works highlight promises of data-driven decision-making in complex network environments. Yet, most prior learning-based systems optimize aggregate network-level objectives, congestion-control parameters, or traffic-engineering policies. This keeps the vacuum of finding outcomes of alternative placements in the network hidden unless the flow is actually moved. \emph{Overall, prior work holds significant promise for improving datacenter congestion management, but the current state of the art still lacks proactive per-flow protection. None of the aforementioned approaches directly solves the problem of deciding whether a currently healthy, performance-sensitive flow should be moved before degradation begins, relying instead on indirect, distributed precursor signals.}


\section{System Model and Problem Formulation}
\label{sec:background}

\subsection{System Model}
\label{sec:system-model}

\begin{figure}[tb]
  \centering
  \includegraphics[width=\columnwidth]{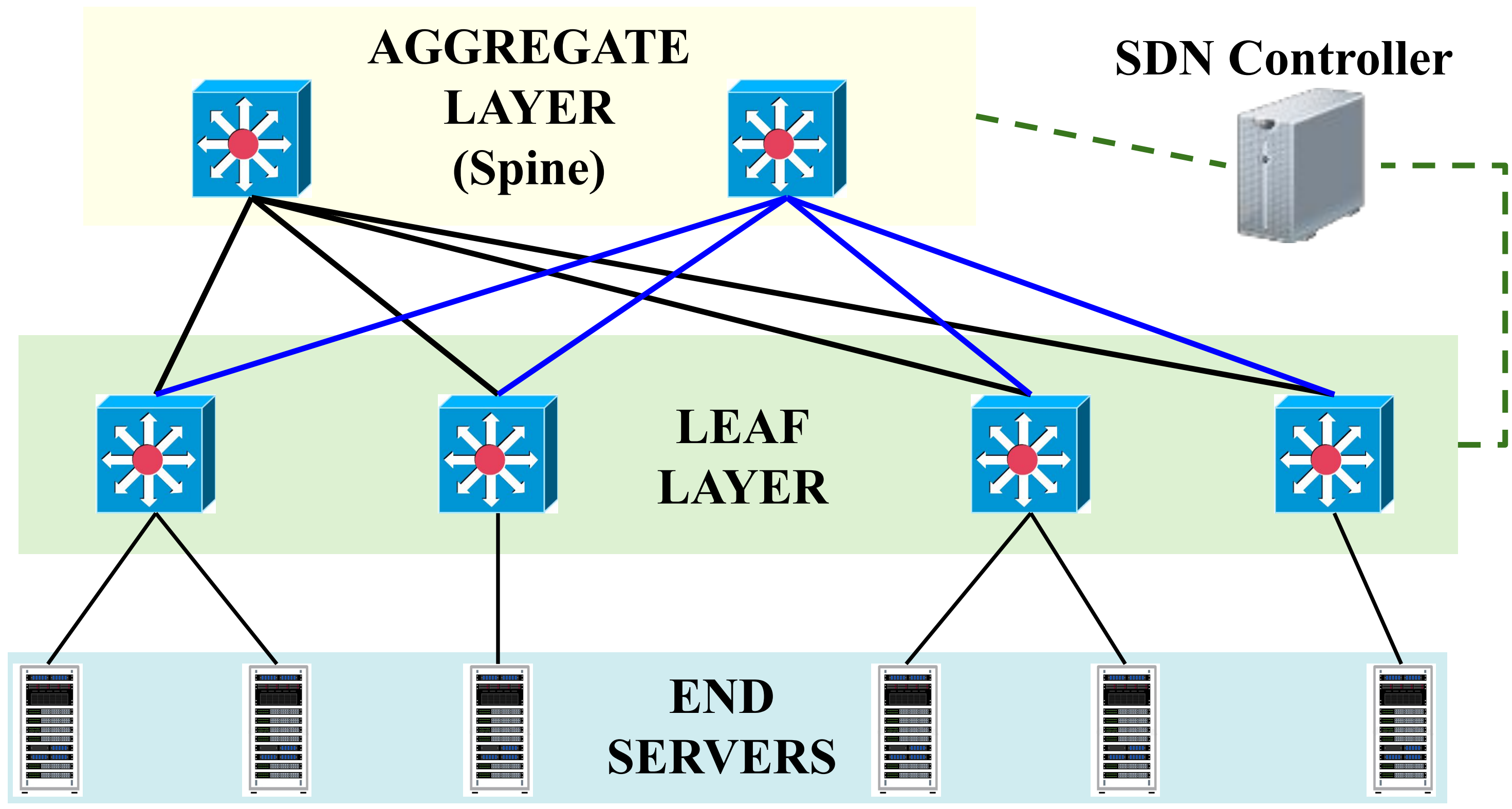}

  \caption{Different layers in a datacenter topology}
  \label{fig:topology}
  \vspace{-0.2in}
\end{figure}

We consider a multi-tenant datacenter network composed of leaf and aggregation switches arranged in a two-tier Clos topology, as illustrated in Figure \ref{fig:topology}. End-hosts attach to leaf switches, and all inter-host traffic is forwarded through one of several aggregation switches, each providing an equal-cost path between source and destination. This architecture offers high path diversity and is widely used in modern datacenter deployments. Path selection in the network is performed using Equal-Cost Multi-Path (ECMP) routing, where each flow is assigned to an aggregation switch based on a hash of its packet header fields. Once assigned, a flow remains on the selected path for its lifetime, as ECMP does not incorporate runtime congestion feedback. As a result, multiple flows may be mapped to the same aggregation switch, leading to potential contention for shared bandwidth.

The network is managed by a centralized SDN controller, as shown in Figure \ref{fig:topology}, which can override default forwarding decisions and reroute individual flows across aggregation switches. 
The controller also collects aggregate telemetry, including traffic rates and activity levels, from the network switches.
These signals provide only a partial view of the network state and do not directly reveal the placement or behavior of individual competing flows.


\subsection{Problem Formulation}

We denote the aggregation-switch set by $\mathcal{S}$, with each switch having uplink capacity $C$, and the set of leaf switches hosting congestion-flow sources by $\mathcal{L}$.
The \emph{protected flow} $f_p$ is a long-lived TCP stream between a fixed source-destination host pair whose throughput must be maintained under multi-tenant load. 
At any time, $f_p$ traverses exactly one aggregation switch, with its current placement denoted by $k^* \in \mathcal{S}$.
A set of $N$ \emph{congestion flows} ${f_1, \ldots, f_N}$, each comprising multiple parallel TCP streams, originates from leaves in $\mathcal{L}$.
The assignment of each congestion flow is fixed at arrival and not directly observable by the controller; only the aggregate effect on each aggregation switch is measurable through telemetry.

At regular intervals the controller collects a telemetry observation vector $\mathbf{o}_t \in \mathbb{R}^{d}$ encoding aggregate signals from $\mathcal{L}$ and $\mathcal{S}$ together with the current placement $k^*$. These signals characterise instantaneous load, deviation from per-episode baselines, and congestion trends. Based on $\mathbf{o}_t$, the controller may issue a reroute command that moves $f_p$ to any $k \in \mathcal{S}$.

Let $\phi(\mathbf{o}_t, k)$ denote the throughput of $f_p$ when placed on aggregation switch $k$ under the network conditions represented by $\mathbf{o}_t$. The proactive aggregation switch selection problem has two tightly coupled objectives. The first is to maximize the cumulative throughput of the protected flow over time:
\begin{equation}
\scriptsize
\max_{k_0, k_1, \ldots, k_T} \; \sum_{t=0}^{T} \phi\!\left(\mathbf{o}_t,\, k_t\right)
\label{eq:objective1}
\end{equation}
subject to the constraint that each decision at time $t$ is made using only the observations available up to that point.

The second objective is to maximize the \emph{lead time} $\tau_{\text{lead}}$, defined as the interval between the controller's reroute decision and the manifestation of congestion on the protected flow's original aggregation switch in the form of throughput degradation of the flow. A positive lead time corresponds to a proactive decision made during the precursor phase, before any observable performance degradation, while a negative lead time corresponds to a reactive response after congestion has already impacted throughput. This scenario is depicted in Figure~\ref{fig:temporal_gap}.

These objectives are generally aligned because earlier rerouting provides more time to reach a less congested destination before throughput degrades. However, positive lead time does not guarantee higher throughput if the selected destination is also congested. We therefore report both metrics and seek placements $k_t \in \mathcal{S}$ that preserve throughput while enabling timely intervention.

A key difficulty is that $\phi(\mathbf{o}_t, k)$ is directly observable only for the current placement $k^*$ — we write the realised throughput compactly as $\phi_t \equiv \phi(\mathbf{o}_t, k^*)$ — while the throughput under alternative placements $k \neq k^*$ remains unknown unless the flow is explicitly rerouted. This introduces a counterfactual decision problem in which the controller must reason about outcomes that cannot be directly measured. Compounding this challenge are the temporal dynamics of congestion. As illustrated in Figure~\ref{fig:temporal_gap}, congestion flows produce early precursor signals in the network shortly after arrival, while the protected flow's throughput remains unaffected for a substantial period. Degradation occurs later, once the token bucket on the congested aggregation switch is exhausted. A reactive strategy that waits for throughput degradation is therefore inherently late.

The controller must therefore use indirect precursor signals in $\mathbf{o}_t$ to predict future congestion and select an aggregation switch for $f_p$ before the delayed throughput consequences become observable.



\section{Solution Strategy}
\label{sec:design}

\subsection{Design Motivation}

The proactive aggregation switch selection problem is predictive and counterfactual, as future congestion must be inferred from current telemetry $\mathbf{o}_t$, and the throughput of $f_p$ under alternative placements $k \neq k^*$ is unobservable unless taken. 
A third requirement is that learning must be performed offline, since exploratory rerouting in a production network would directly degrade the protected flow's performance. 
These constraints motivate an offline learning formulation that infers future congestion from collected telemetry while reasoning about unobserved outcomes.
\emph{ProFlow} 
adopts this formulation to learn a proactive placement policy entirely offline, which is then deployed as a fixed inference component within the controller.

\subsection{ProFlow Overview}

We hereby discuss \emph{ProFlow}, a routing mechanism implemented within the SDN controller that enables proactive placement of performance-sensitive flows in multi-tenant datacenter networks. Building on the requirements outlined above, the key idea is to leverage real-time network telemetry to anticipate congestion and adjust flow placement before performance degradation occurs, rather than reacting to it afterwards.

At runtime, the system operates in a closed loop. At each decision interval, the controller collects a telemetry observation $\mathbf{o}_t$ that summarises current network conditions across aggregation switches in $\mathcal{S}$ and leaf switches in $\mathcal{L}$. Based on this observation, \emph{ProFlow} evaluates the expected future performance of the protected flow under different placement options and selects an aggregation switch that is likely to remain uncongested. If the selected switch differs from the current placement $k^*$ and satisfies a stability condition, the controller issues a reroute command through the SDN control plane.

The system consists of three logical components. A \emph{signal collection} module gathers network telemetry from switches and constructs the observation vector $\mathbf{o}_t$ used for decision making. A \emph{decision module} processes this observation to evaluate candidate placements and determine whether a reroute should be performed. Finally, the \emph{SDN controller} enforces the selected placement by updating forwarding rules for $f_p$.
To this end, \emph{ProFlow} employs an offline-trained model that captures the relationship between observed telemetry signals and future congestion, as well as the expected impact of placement decisions. This model is trained on data collected from curated congestion scenarios on a physical testbed, and deployed as a fixed inference component that processes real-time telemetry without further learning or exploration.

\subsection{State Representation}

At each decision interval, ProFlow converts the raw telemetry observation $\mathbf{o}_t$ into a structured state representation $\mathbf{s}_t \in \mathbb{R}^{d_s}$ used for decision making. The state vector is constructed by grouping and normalising telemetry signals into four components:
\begin{equation}
\scriptsize
\mathbf{s}_t = \big[\, \mathbf{x}^{\text{agg}}_t,\; \mathbf{x}^{\text{leaf}}_t,\; \mathbf{h}_t,\; \mathbf{x}^{\text{flow}}_t \,\big]
\label{eq:state_decomposition}
\end{equation}

The first component, $\mathbf{x}^{\text{agg}}_t \in \mathbb{R}^{5|\mathcal{S}|}$, summarises per-aggregation-switch congestion state. For the current host $k^*$, overflow and flow-count signals are encoded as \emph{signed} deviations from per-episode baselines, exposing both the protected flow's own TCP collapse (an early sign of severe congestion on the current path) and crowd accumulation (the primary signal used for proactive rerouting); for all other switches $k \neq k^*$, unsigned normalisations are used. The second component, $\mathbf{x}^{\text{leaf}}_t \in \mathbb{R}^{2|\mathcal{L}|}$, captures precursor signals at the leaf switches in $\mathcal{L}$. The third component, $\mathbf{h}_t \in \{0,1\}^{|\mathcal{S}|}$, is a one-hot encoding of the current placement $k^*$. The fourth component, $\mathbf{x}^{\text{flow}}_t \in \mathbb{R}^{4}$, captures protected-flow throughput $\phi_t$, aggregate congestion-flow intensity, and short-window degradation signals that allow the controller to detect throughput drop on the current path before it becomes severe. The full per-coordinate signal definitions, normalisation, scaling constants $C_x$, and signed/unsigned encoding rules are given in Appendix~\ref{app:state}.

\begin{figure*}[!t]
\centering
\includegraphics[width=0.8\textwidth]{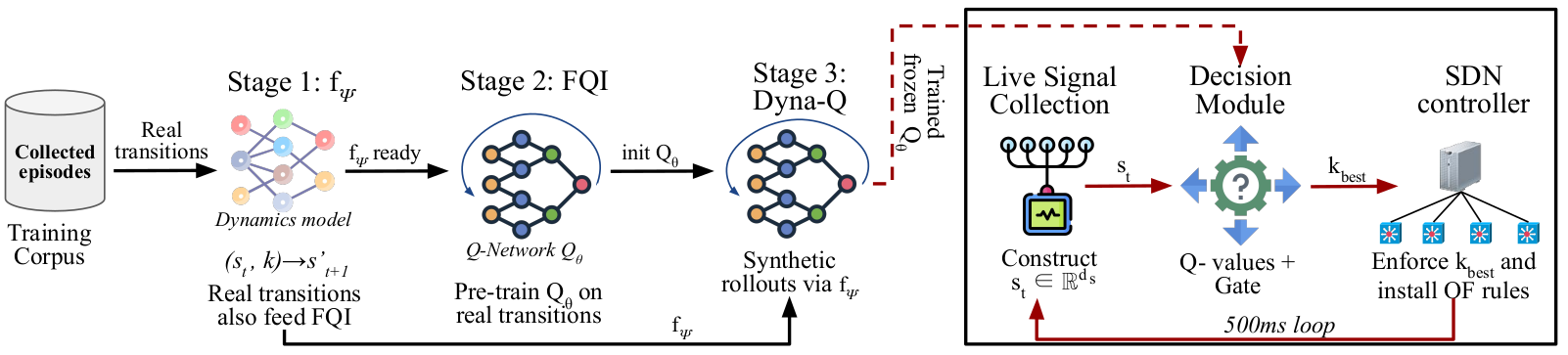}
\vspace{-0.05in}
\caption{ProFlow pipeline. \textit{Offline (left):} a dynamics model $f_\psi$ is trained on offline episodes collected from controlled congestion scenarios (Stage~1); the Q-network $Q_\theta$ is then pre-trained on real transitions via fitted Q-iteration (Stage~2) and refined with Dyna-style synthetic rollouts
generated by $f_\psi$ (Stage~3). \textit{Online (right):} the trained $Q_\theta$ is deployed as a frozen inference component within the SDN controller. At each 500\,ms decision interval, telemetry from leaf and aggregation switches is normalised into the state vector $\mathbf{s}_t \in \mathbb{R}^{d_s}$, the decision module computes Q-values and applies a stability gate (margin $\delta$, $W$ consecutive votes, cooldown) to select $k_{\text{best}}$, and the SDN controller installs per-flow
OpenFlow rules to enforce the chosen aggregation switch.}
\label{fig:pipeline}
\vspace{-0.2in}
\end{figure*}

\subsection{Action Space and Rerouting Mechanism}

ProFlow models the placement decision for the protected flow as a discrete action selection problem. At each decision interval, the controller selects an aggregation switch $k \in \mathcal{A}$, where the action set $\mathcal{A} = \mathcal{S}$ identifies each action with placing the protected flow on a specific aggregation switch. Once an action is selected, the SDN controller enforces the corresponding placement by installing per-flow forwarding rules on the relevant leaf switches. The update is applied to both the forward path and the reverse path to ensure consistent routing.

A key challenge in this setting is avoiding excessive rerouting. During the precursor phase, congestion indicators may fluctuate across aggregation switches, leading to small variations in the estimated quality of different placement options. If the controller always selects the best-scoring switch at each interval, this can result in frequent switching between aggregation switches, which may disrupt the transport layer and degrade overall performance.
To address this, ProFlow employs a three-part stability mechanism. Let $Q(\mathbf{s}_t, k)$ denote the estimated value of placing the protected flow on aggregation switch $k$ under state $\mathbf{s}_t$. At each step, the controller computes the highest-scoring action $k_{\text{best}} = \arg\max_{k \in \mathcal{A}} Q(\mathbf{s}_t, k)$ and the Q-value gap $\Delta Q = Q(\mathbf{s}_t, k_{\text{best}}) - Q(\mathbf{s}_t, k^*)$.

First, a reroute candidate is admitted only if $\Delta Q > \delta$, where $\delta$ is a margin threshold that filters noise-driven switches. Second, the same candidate switch must win for $W$ consecutive decision steps before a reroute is issued, ensuring that the controller acts on a consistent signal rather than a transient fluctuation. Third, once a reroute has been issued, a cooldown period suppresses further reroutes, allowing the transport layer to stabilise before the next decision. A reroute is triggered only when all three conditions are simultaneously satisfied.

\subsection{Model-Based Offline Reinforcement Learning}

ProFlow formulates the placement problem as a reinforcement learning problem and learns a value function from offline data. At each decision interval, the controller observes the current state $\mathbf{s}_t$, selects an action $k \in \mathcal{A}$, and receives a scalar reward $r_t$ that captures how well the chosen placement served the protected flow. The goal is to learn a function $Q(\mathbf{s}, k)$ that estimates the expected long-term return of choosing switch $k$ in state $\mathbf{s}$:

\begin{equation}
\scriptsize
Q(\mathbf{s}_t, k) = \mathbb{E} \left[ \sum_{i=0}^{\infty} \gamma^i \, r_{t+i} \,\middle|\, \mathbf{s}_t, k_t = k \right],
\end{equation}

where $\gamma \in (0,1)$ is a discount factor. At runtime, the controller selects the action with the highest Q-value, subject to the stability gate. The Q-function is implemented as a multi-layer feedforward neural network $Q_\theta : \mathbb{R}^{d_s} \rightarrow \mathbb{R}^{|\mathcal{A}|}$ that maps the state vector directly to a vector of values, one per candidate action.

\paragraph{Reward design}
The reward jointly encodes the two objectives from Section~\ref{sec:background}: high throughput on the current placement, and early action driven by precursor signals.

For transitions observed in the collected data, the reward is defined as:

\begin{equation}
\scriptsize
\begin{aligned}
r_t = \mathrm{clip}\!\Big(&
\phi_{\text{norm}}
- \alpha_{\text{tc}}\,\hat{\xi}_{k^*}
+ \alpha_{\text{stay}}\,(1 - \hat{\xi}_{k^*}) \\
&- \alpha_{\text{coll}}\,(1 - \phi_{\text{norm}})
+ \beta\,\Delta n_{k^*}^{+},
\; -1,\; 2
\Big)
\end{aligned}
\label{eq:real_reward}
\end{equation}

where $\phi_{\text{norm}} = \phi_{t+1} / \phi_0$ is the next-step throughput normalised by the per-episode pre-congestion baseline $\phi_0$, and $\hat{\xi}_{k^*} = \xi_{k^*} / C_\xi$ is the normalised token-bucket overflow rate on the current host $k^*$. The term $\Delta n_{k^*}^{+} = \max(0,\, (n_{k^*} - \bar{n})/C_n)$ is the positive excess flow count on $k^*$, where $\bar{n}$ and $C_n$ are defined in Appendix~\ref{app:state}. The clipping bounds prevent extreme Q-value targets during training.

The term $\phi_{\text{norm}}$ rewards high throughput. The terms weighted by $\alpha_{\text{tc}}$ and $\alpha_{\text{stay}}$ 
together penalise placement on a congested aggregation switch and reward placement on a clean one. The term weighted by $\alpha_{\text{coll}}$ additionally penalises states where the protected flow's throughput has already collapsed below baseline. These terms are all grounded in signals that are directly observable in the current state.

The crowd term $\beta\,\Delta n_{k^*}^{+}$ ($\beta < 0$) is the proactive component: it fires as soon as excess flows arrive on $k^*$, before the token bucket is exhausted and before throughput degrades. By penalising crowd accumulation independently of throughput, this term directly incentivises rerouting during the precursor window, which is the core behavioural objective of the system.

\paragraph{Handling counterfactual outcomes with a learned dynamics model}
Because the data-collection policy did not freely move $f_p$ to all switches at all times, the Q-function must evaluate placements rarely seen in the data and reason about future states that depend on actions not yet executed. To address this, ProFlow learns a dynamics model $f_\psi$ in addition to the Q-function. The dynamics model is a feedforward neural network that takes as input the current state $\mathbf{s}_t$ and a one-hot encoding of the selected action $k$, and predicts the resulting next state $\mathbf{s}'_{t+1}$. Given this model, the agent can simulate the consequences of any placement decision from any state, even those not present in the collected data.

\paragraph{Reward for synthetic transitions}
Because the dynamics model is used to generate rollouts under hypothetical placement decisions, particularly reroutes that were not executed in the data, the reward applied to these synthetic transitions is designed to be more explicitly proactive than the real-transition reward. For a reroute action, the synthetic reward augments the real-transition reward's throughput/congestion structure with three additional shaping terms: a destination-congestion penalty on the chosen switch, a proactive-escape bonus that rewards moving from a more congested switch to a cleaner one, and a lead-time bonus that rewards rerouting while $k^*$ is still clean, explicitly incentivising early action. For a stay action, analogous terms reward remaining on a clean switch while penalising congestion and throughput collapse, including a crowd penalty consistent with the real-transition reward. The full form is given in Appendix~\ref{app:reward}.

\paragraph{Training procedure}
Training proceeds in three offline stages.
In the first stage, the dynamics model $f_\psi$ is trained on transitions extracted around observed reroute events, learning to predict how the network state evolves following a placement change. This model is the foundation for all subsequent planning.
In the second stage, the Q-network $Q_\theta$ is pre-trained on all real transitions using fitted Q-iteration (FQI). This stage provides a stable initialisation of the Q-function by regressing Q-values toward Bellman targets computed from the real-transition reward and bootstrapped next-state values. In the third stage, the Q-network is refined using a Dyna-style approach. For each batch of real transitions, the agent also performs a series of synthetic rollout steps using $f_\psi$, with rewards computed by the synthetic reward function. These rollouts allow the agent to evaluate placement choices that were not taken in the data — in particular, early reroutes during the precursor phase — enabling it to learn proactive behaviour that the real data alone cannot fully supervise. The entire pipeline for \emph{ProFlow} is shown in Figure~\ref{fig:pipeline}.



\section{Evaluation and Results}

\label{sec:eval-results}

\subsection{Implementation}
\label{sec:implementation}

\paragraph{Testbed}
ProFlow is implemented on the NSF FABRIC testbed using a leaf-aggregation switch topology composed of Open vSwitch instances. Four aggregation switches and four leaf switches are deployed on dedicated compute nodes, each running OVS; three of the four leaves (constituting $\mathcal{L}$) host the congestion-source hosts. Eight end-hosts are attached to the leaf layer, two per leaf. A centralized Ryu SDN controller manages all switches and hosts the signal collection module, the Q-function inference engine, and the rerouting logic. Egress rate limiting on each aggregation switch uplink is enforced using Linux \texttt{tc} token-bucket filters.

\paragraph{Signal collection and telemetry}
The controller polls port statistics and flow-level byte counters from all switches via OpenFlow at 500\,ms intervals. Token-bucket overflow rates are collected concurrently from each aggregation switch via a lightweight HTTP endpoint running on each node. These raw measurements are combined to construct the state vector $\mathbf{s}_t$ as described in Section~\ref{sec:design}.

\paragraph{Training data and setup}

The Q-network $Q_\theta$ and dynamics model $f_\psi$ are trained entirely offline using controlled testbed episodes generated under the strategies listed in Table~\ref{tab:training_strategies} (Appendix~\ref{app:strategies}). Training is performed in PyTorch on a CUDA-capable GPU and completes in under ten minutes.

\paragraph{Evaluation scenarios}
The agent is evaluated across ten main scenarios (S1--S10) and two ablation scenarios (C1, C2), all run on the physical testbed. Each scenario is repeated three times under three policies (agent, reactive, static (no reroute)), yielding $10 \times 3 \times 3 = 90$ main-test episodes and $2 \times 3 = 6$ ablation episodes per model variant. Table~\ref{tab:scenarios} (Appendix~\ref{app:scenarios}) summarises each scenario.

\paragraph{Baseline thresholds}
The reactive baseline reroutes when the token-bucket overflow rate on the protected flow's current switch exceeds $\xi_{\text{th}} = 27{,}000$ bytes/s for three consecutive samples, chosen as the lowest threshold that never false-alarms on the protected flow's own overflow, as detailed in Appendix~\ref{app:threshold}. Similarly, the crowd signal reroutes when the flow count on the protected flow's switch exceeds a threshold, set to $T = 5$, chosen to sit between the flow count produced by a single congester and the smallest genuine multi-host crowd, as detailed in
 Appendix~\ref{app:crowd}.


\subsection{Results}
\label{sec:results}

We evaluate ProFlow against two baselines, static and reactive (Section~\ref{sec:implementation}), across the ten main scenarios and two crowd-only ablation scenarios defined in Appendix~\ref{app:scenarios}. We first present per-scenario throughput timeseries, then aggregate mean throughput across scenarios. We then quantify the lead-time advantage of ProFlow over reactive and contrast their token-bucket overflow profiles. A reward-signal sweep and crowd-signal ablation isolate the contribution of each design choice. We close with a seed-robustness study demonstrating that ProFlow's behaviour is reproducible across training seeds.


\paragraph{Per-Scenario Throughput Timeseries}
\label{ssec:timeseries}

\begin{figure*}[t]
    \centering

    \includegraphics[width=0.35\textwidth]{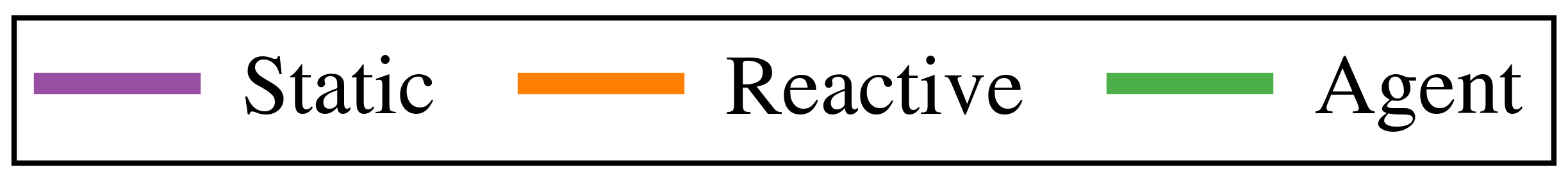}

    \setlength{\tabcolsep}{1pt}
    \begin{tabular}{@{}ccccc@{}}
        \includegraphics[width=0.195\textwidth]{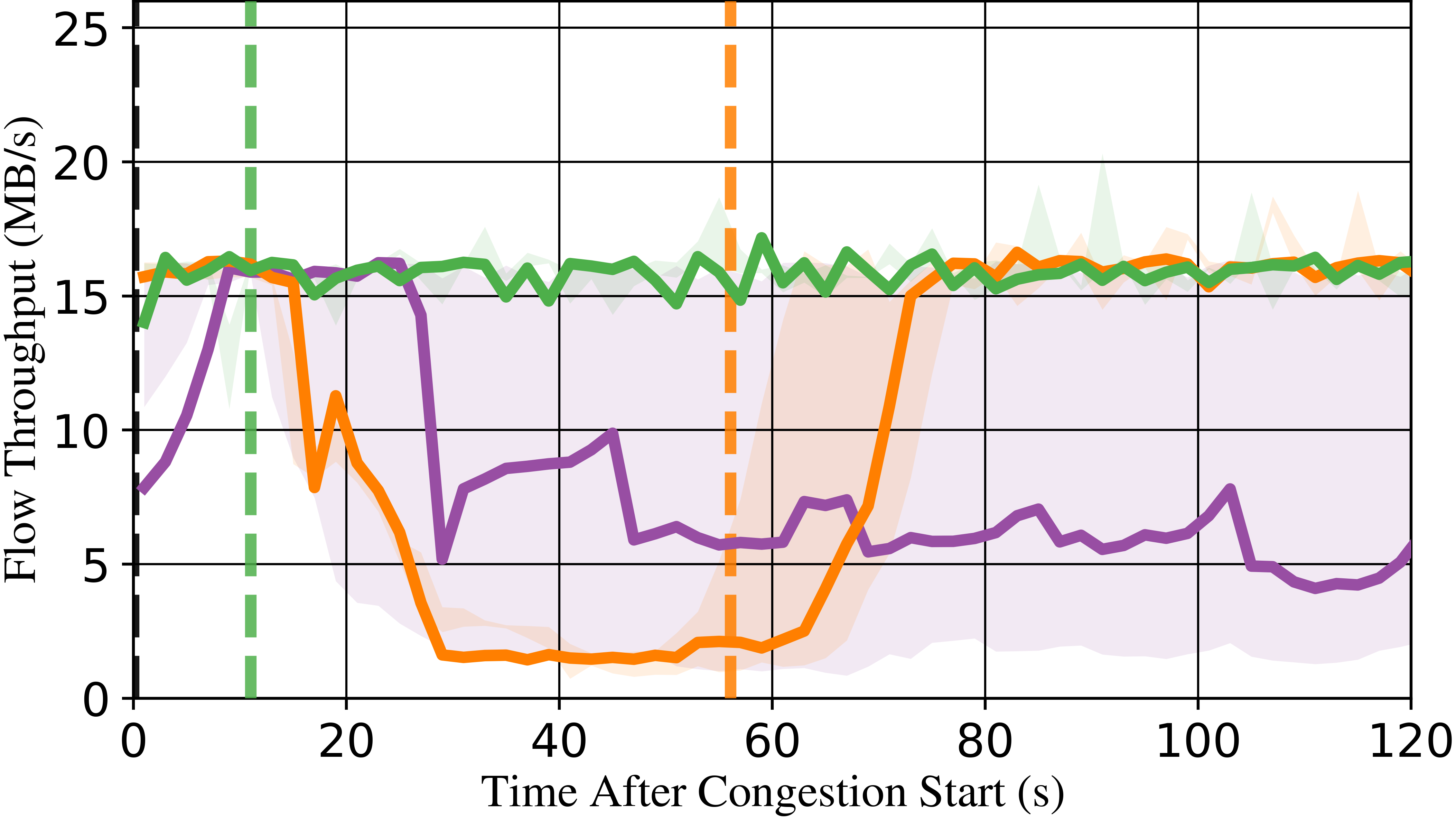} &
        \includegraphics[width=0.195\textwidth]{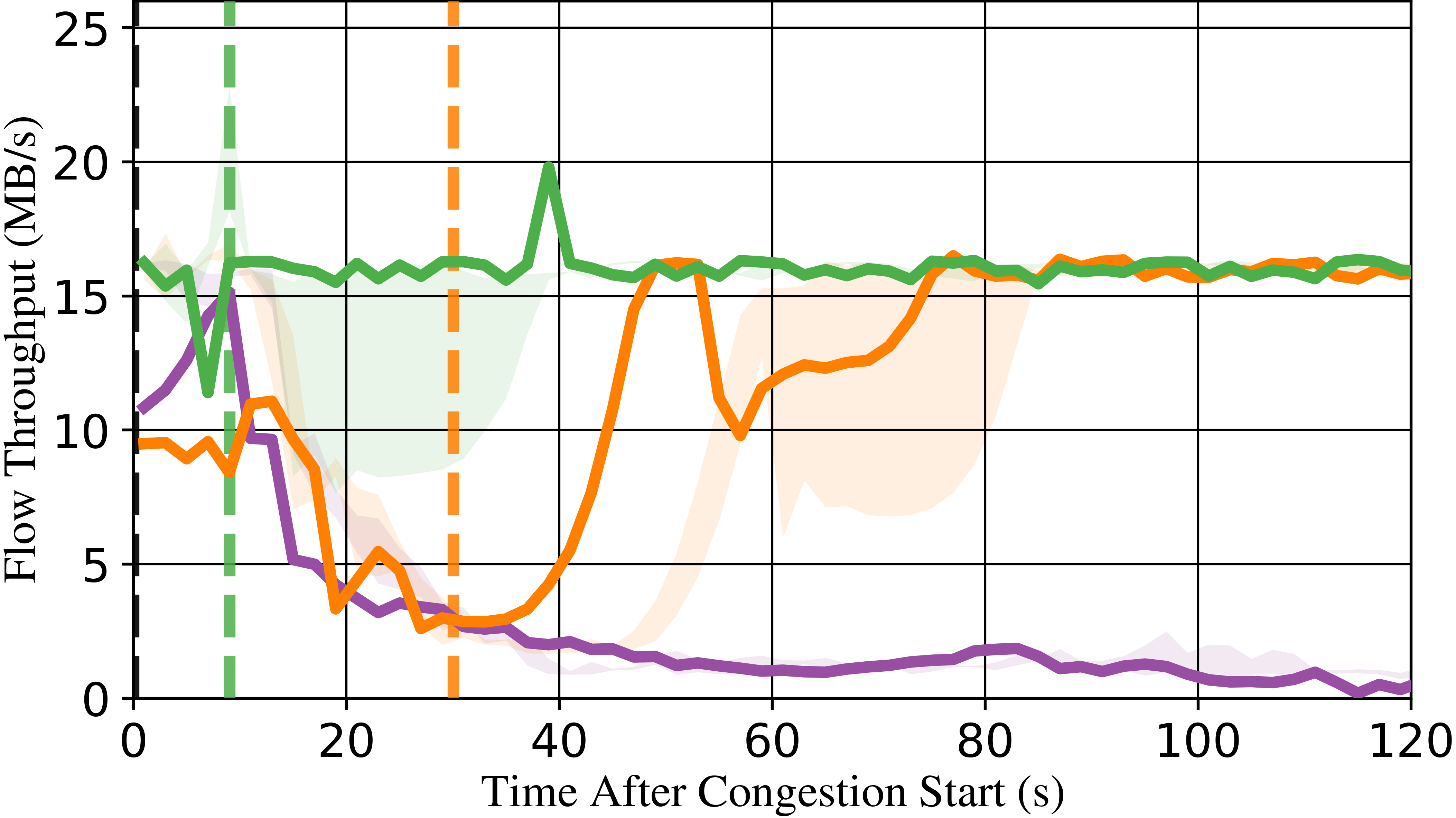} &
        \includegraphics[width=0.195\textwidth]{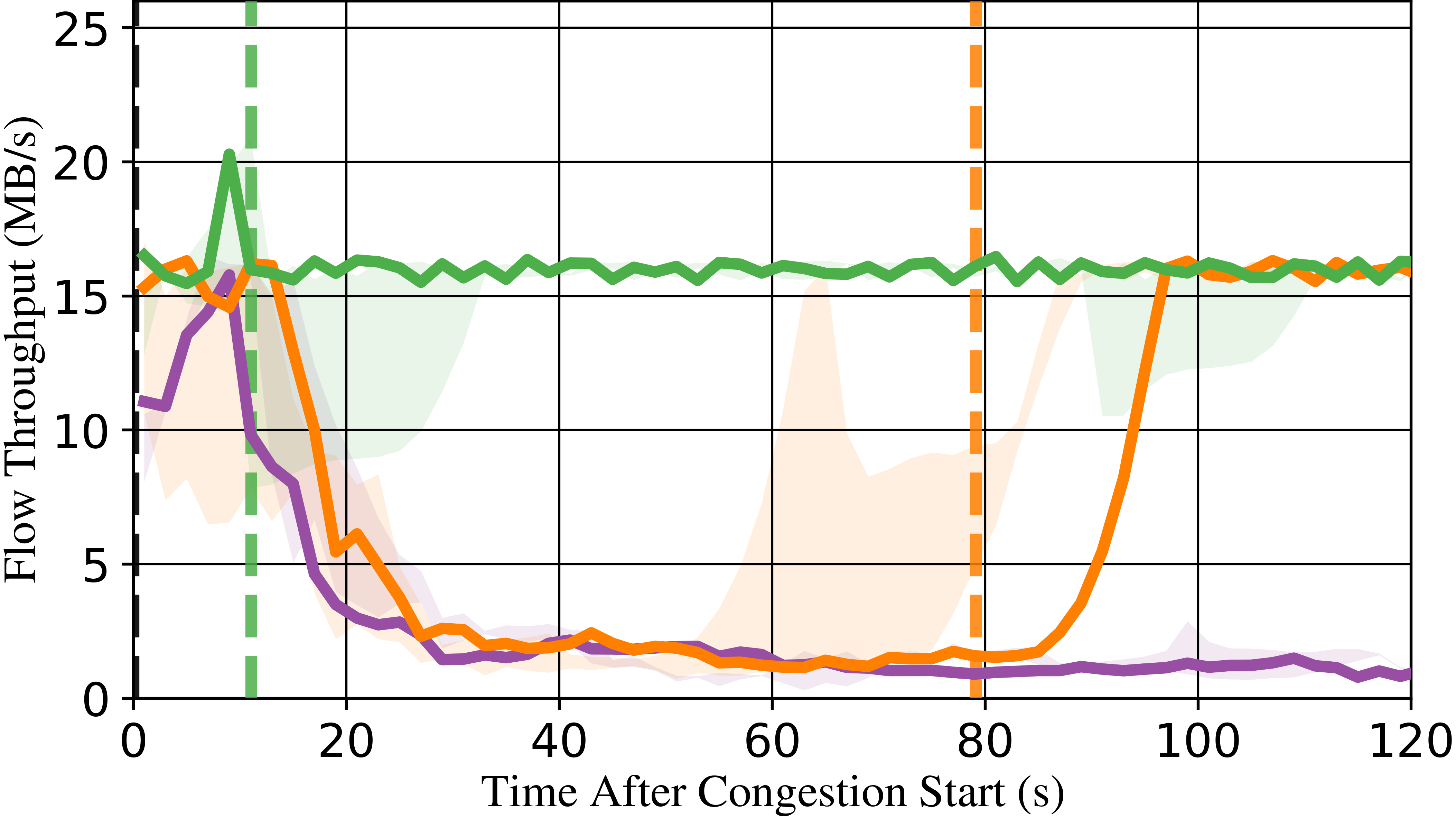} &
        \includegraphics[width=0.195\textwidth]{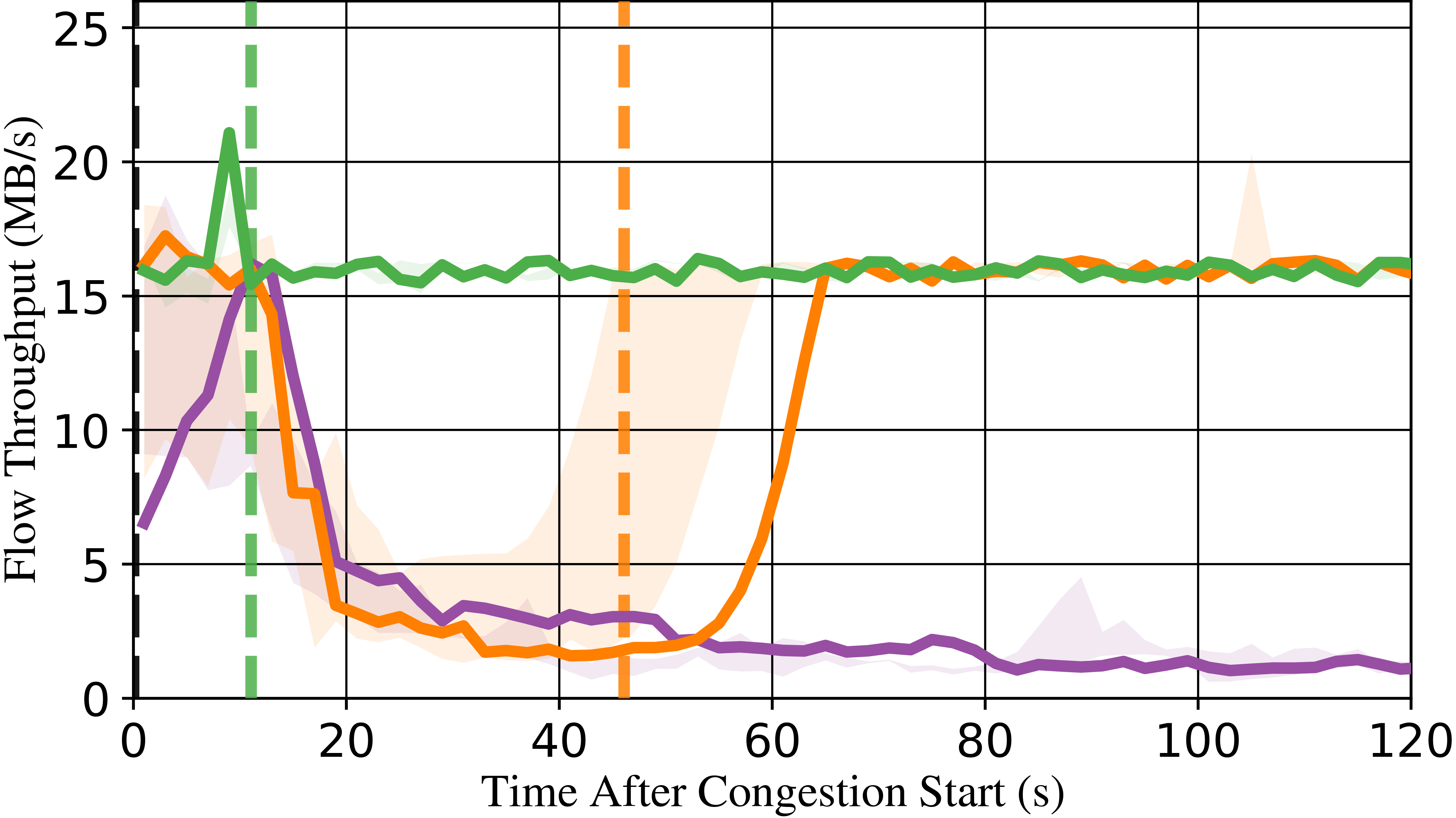} &
        \includegraphics[width=0.195\textwidth]{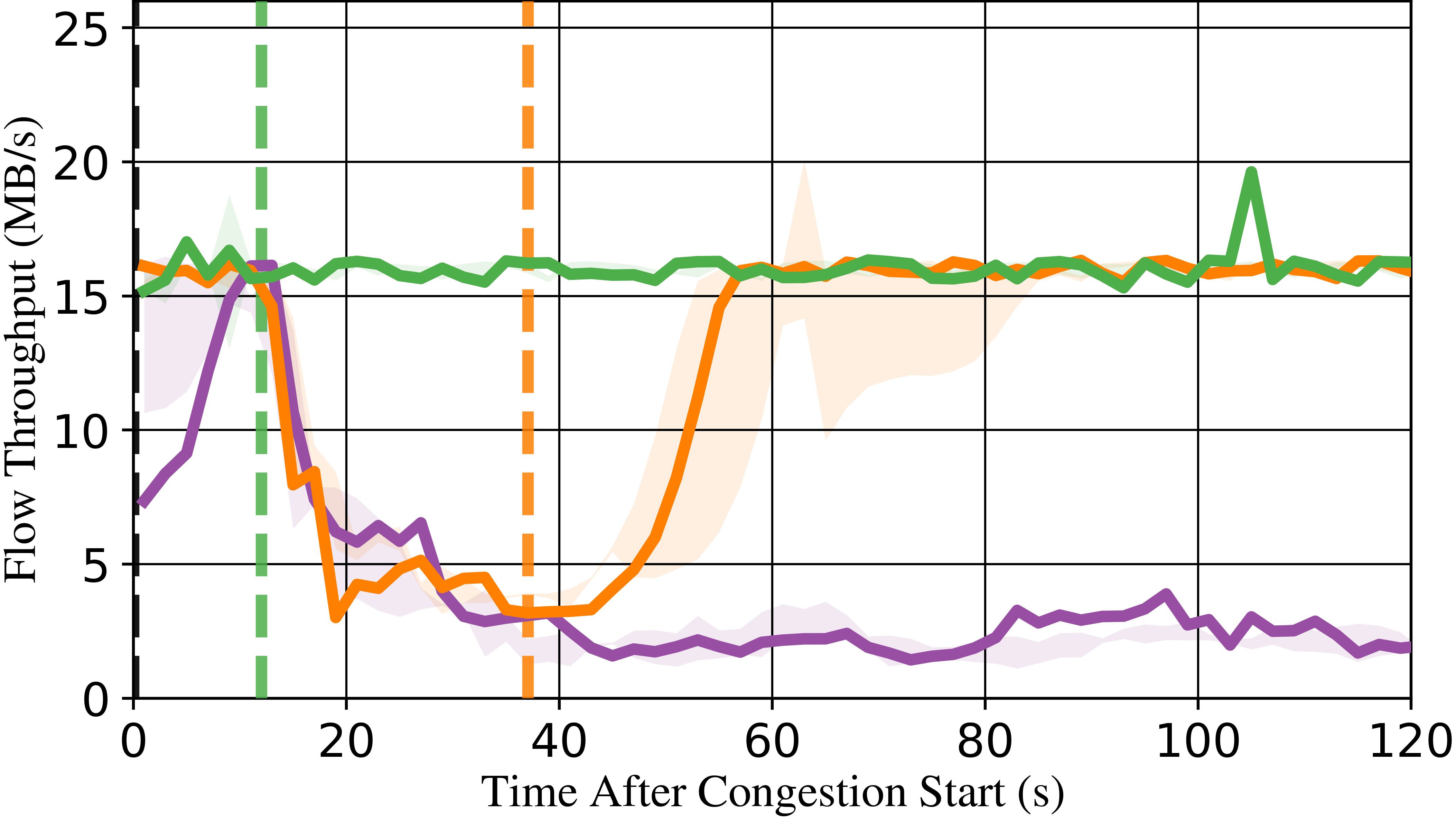} \\[-2pt]
        \small S1 & \small S2 & \small S3 & \small S4 & \small S5 \\[3pt]

        \includegraphics[width=0.195\textwidth]{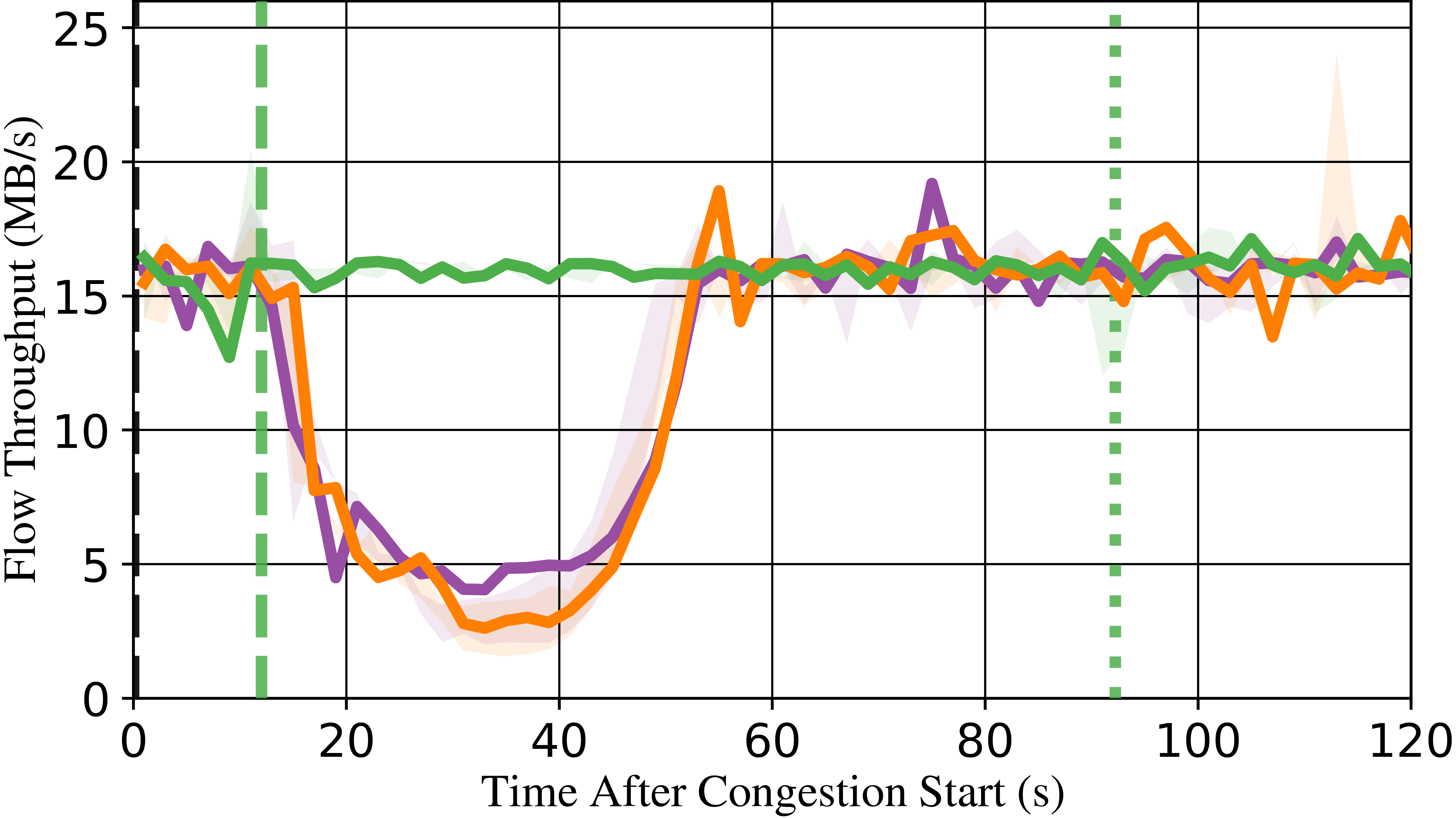} &
        \includegraphics[width=0.195\textwidth]{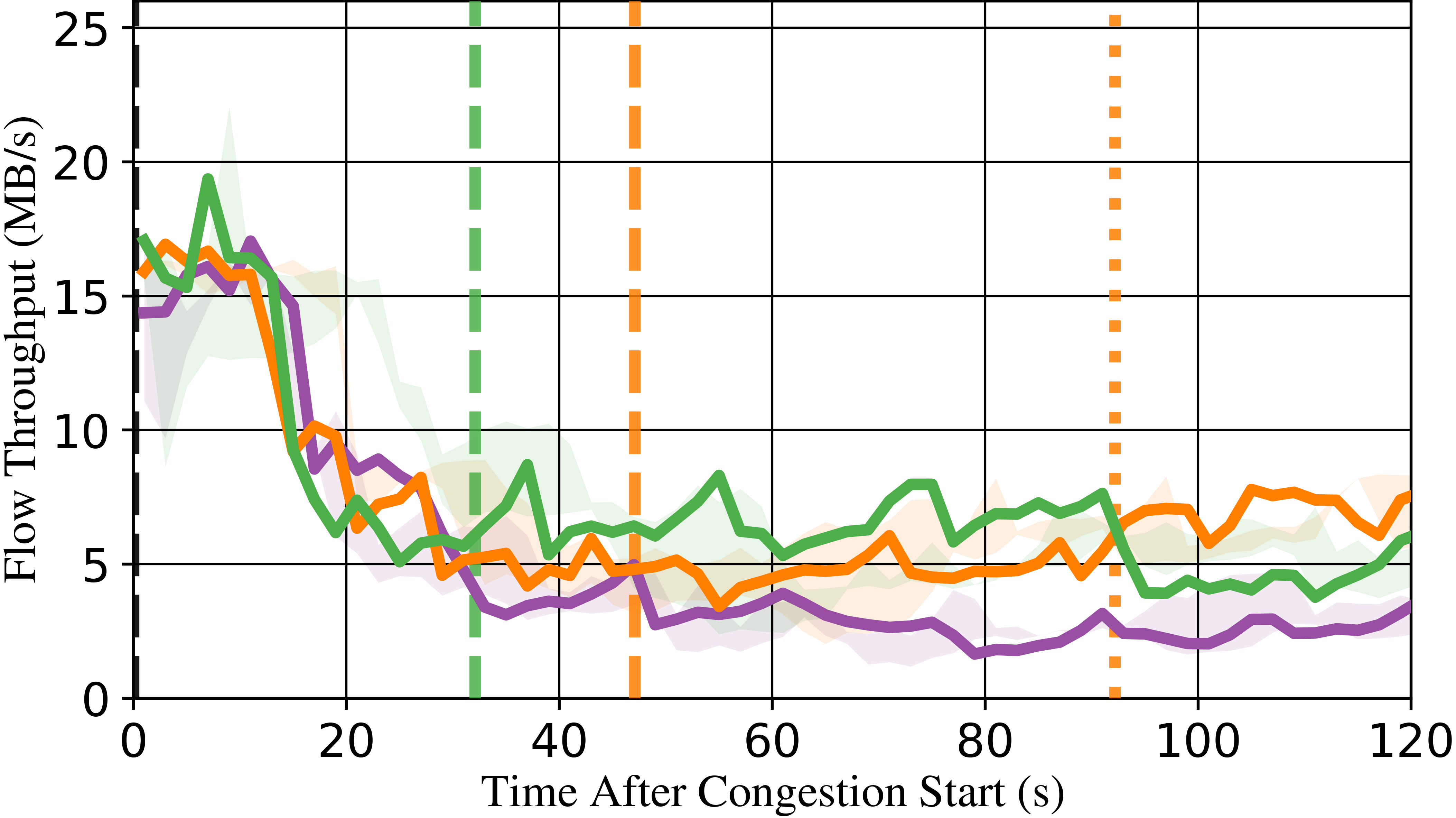} &
        \includegraphics[width=0.195\textwidth]{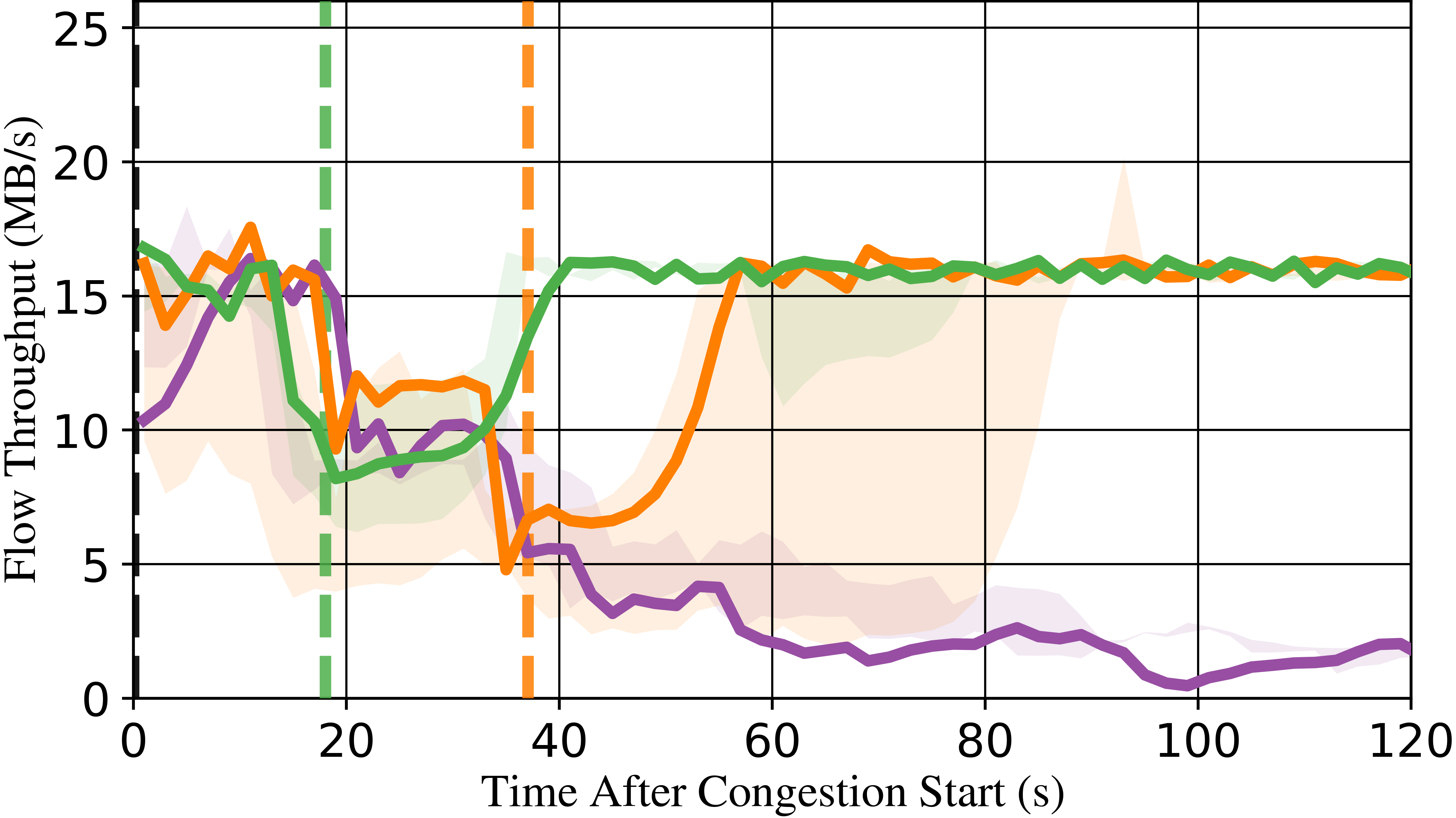} &
        \includegraphics[width=0.195\textwidth]{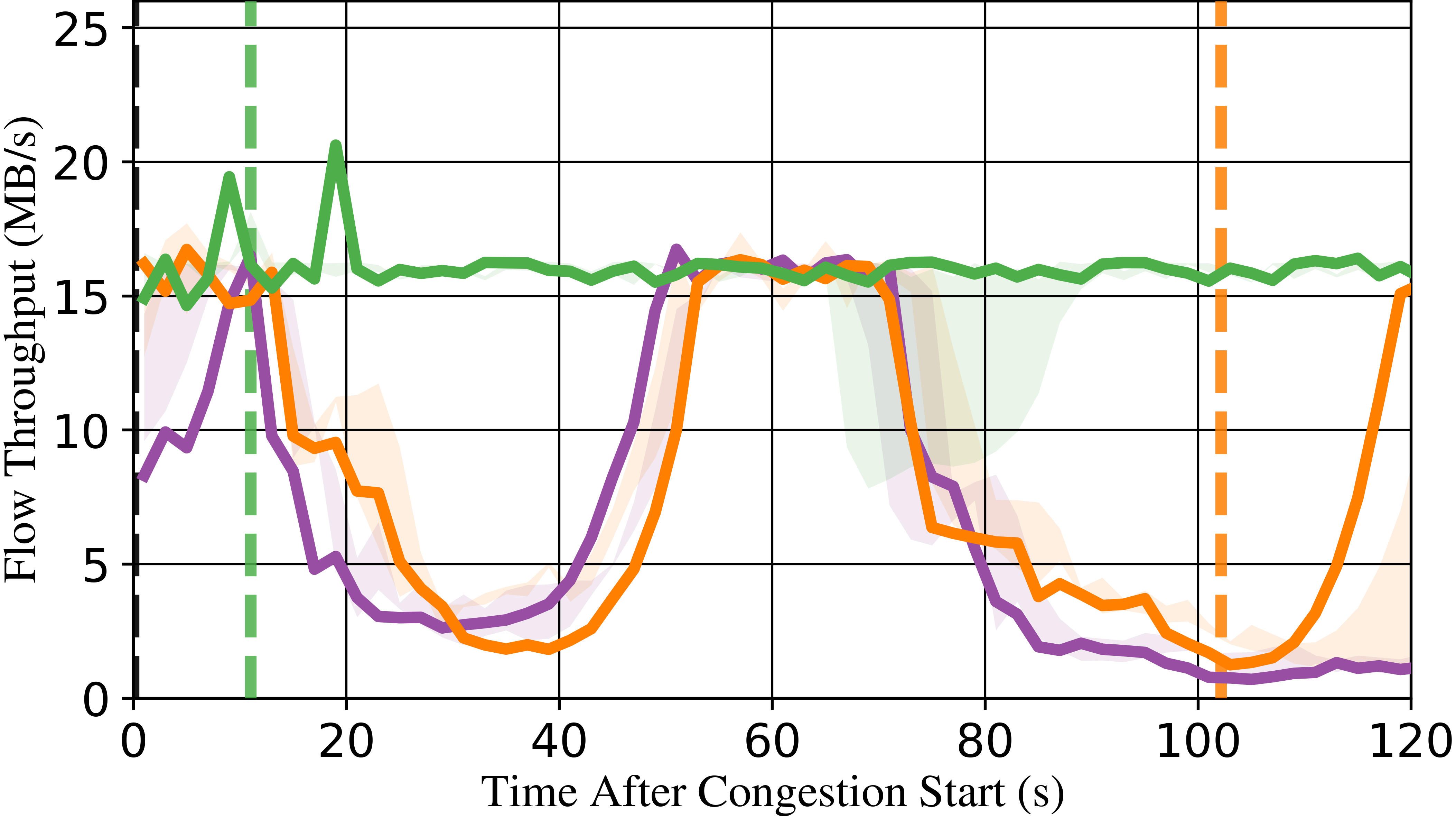} &
        \includegraphics[width=0.195\textwidth]{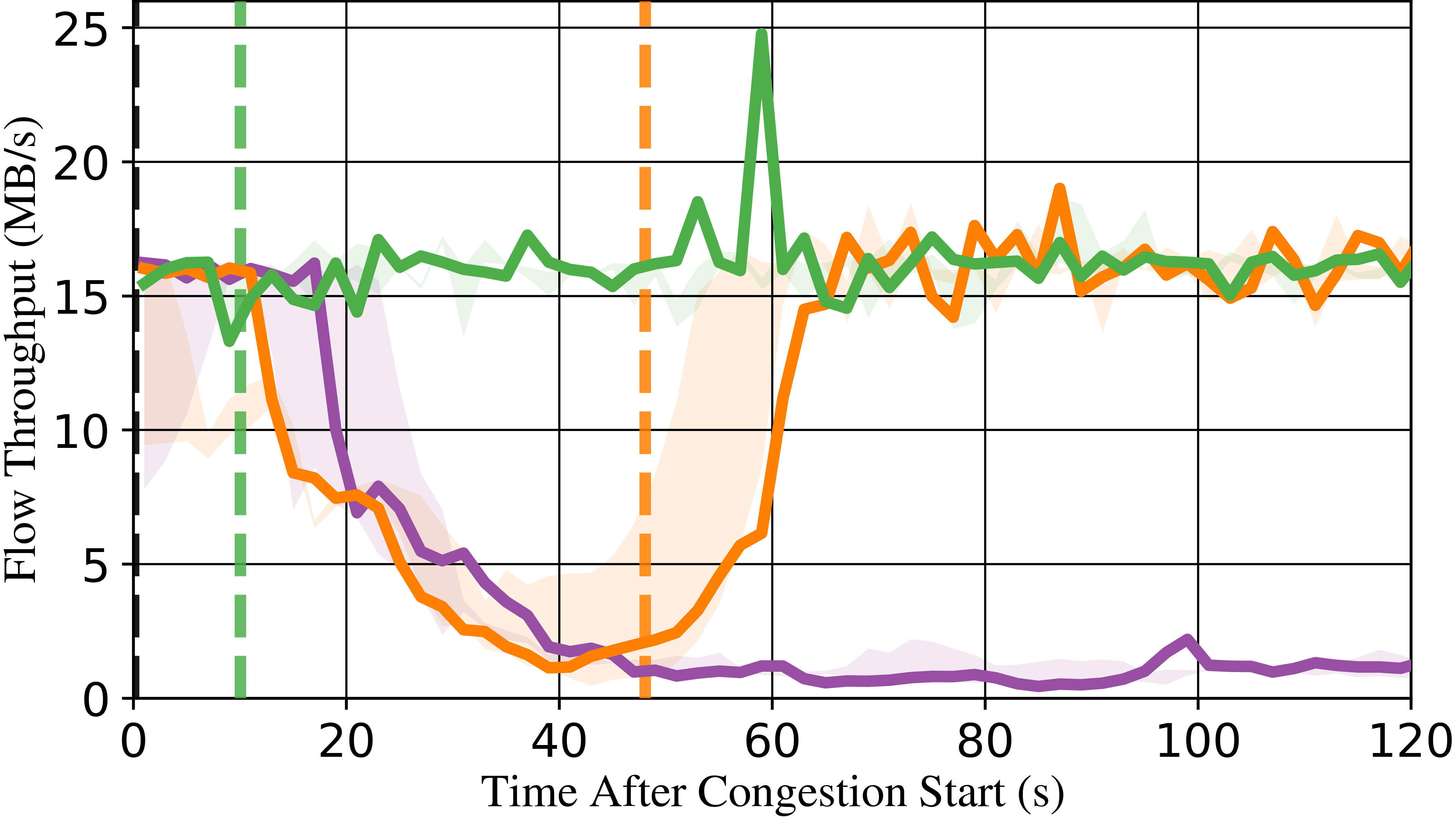} \\[-2pt]
        \small S6 & \small S7 & \small S8 & \small S9 & \small S10
    \end{tabular}

    \vspace{-0.5em}

    \caption{Throughput of $f_p$ (MB/s) over the 120\,s congestion window for all ten evaluation scenarios. Each panel shows the median episode per policy (Static: solid purple; Reactive: solid orange; Agent: solid green), with shaded bands indicating the min-max range across the remaining two repeats. Vertical dashed and dotted markers indicate first and second reroute times of the median episode, respectively.}
    \label{fig:timeseries_all}
    \vspace{-0.2in}
\end{figure*}

Figure~\ref{fig:timeseries_all} shows the throughput of $f_p$ over the 120\,s congestion window for all ten scenarios. Each panel overlays three policies: Static (solid purple), Reactive (solid orange), and Agent (solid green). Lines show the median episode per policy, selected by mean throughput across three repeats. Shaded bands show the min-max range of the remaining two repeats, reflecting run-to-run variability without averaging artefacts. Vertical markers indicate the first reroute time of the median episode (dashed) and a second reroute where applicable (dotted).

Across most scenarios, the agent exhibits a consistent behavioural pattern. It detects the crowd signal $\Delta n_{k^*}^{+}$ within 10--18\,s of congestion start and reroutes while $f_p$ is still operating near its healthy baseline. The flow experiences little to no throughput degradation, and the shaded bands are narrow, indicating that this behaviour is reproducible across repeats. The static policy sustains near-zero throughput throughout once congestion arrives, as $f_p$ competes indefinitely with congestion traffic on the same switch. The reactive policy recovers, but only after throughput has already collapsed. $\xi_{k^*}$ must exceed $\xi_{\text{th}}$ for three consecutive seconds before a reroute fires, by which point $f_p$ has degraded substantially. This delay is visible in every panel as a prolonged low-throughput period before the reactive reroute marker. In S6, the reactive policy never fires at all because the rolling congestion pattern clears each switch before $\xi_{\text{th}}$ can be sustained, making reactive functionally indistinguishable from static. In S9, the burst pattern similarly defeats reactive. The 30\,s burst window ends before three consecutive seconds above $\xi_{\text{th}}$ are accumulated, and by the time the second burst triggers a reroute, only 18\,s of the episode remain.

Two scenarios depart from this dominant pattern and merit closer examination. S7 is the only scenario where the agent does not outperform reactive: two congesters share $k^*$ and three are placed on a separate switch, leaving no fully clean destination at decision time. The agent correctly detects the crowd signal and reroutes, but selects a destination switch already carrying three congestion flows, trading one partially congested placement for a more heavily congested one. Reactive ends up at a similarly congested destination in its median episode, but issues a second reroute after the cooldown expires and recovers marginally better, leaving a 0.06\,MB/s margin, effectively a tie. The failure here is behavioral rather than structural. The state already carries the information needed to rank destinations, and the agent does use it elsewhere, correctly avoiding a congested destination, as shown in Appendix~\ref{app:crowd_scenarios}. S7 is harder because every candidate switch is already congested, so the differences are small and the policy does not reliably prefer the marginally cleaner one, a training-coverage limit rather than a representational one. S8 differs in the opposite way: congesting flows arrive sequentially, one every 25\,s, so the crowd signal builds gradually rather than saturating, and the agent issues its reroute around 18\,s, later than in the full-overlap scenarios. The agent still beats reactive by a substantial margin, but a transient dip is visible in the agent line immediately after rerouting, caused by TCP slow-start on the new (clean) path as the connection must renegotiate its congestion window before recovering to peak throughput, which it does within 10-15\,s. The wider shaded band in S8 relative to the full-overlap scenarios reflects the timing sensitivity of the ramp-up pattern where small differences in when the last of all the consecutive vote accumulates determine how much of the slow-start dip falls inside the measurement window.

\paragraph{Mean Flow Throughput}

\begin{figure}[t]
\centering
\includegraphics[width=0.9\columnwidth]{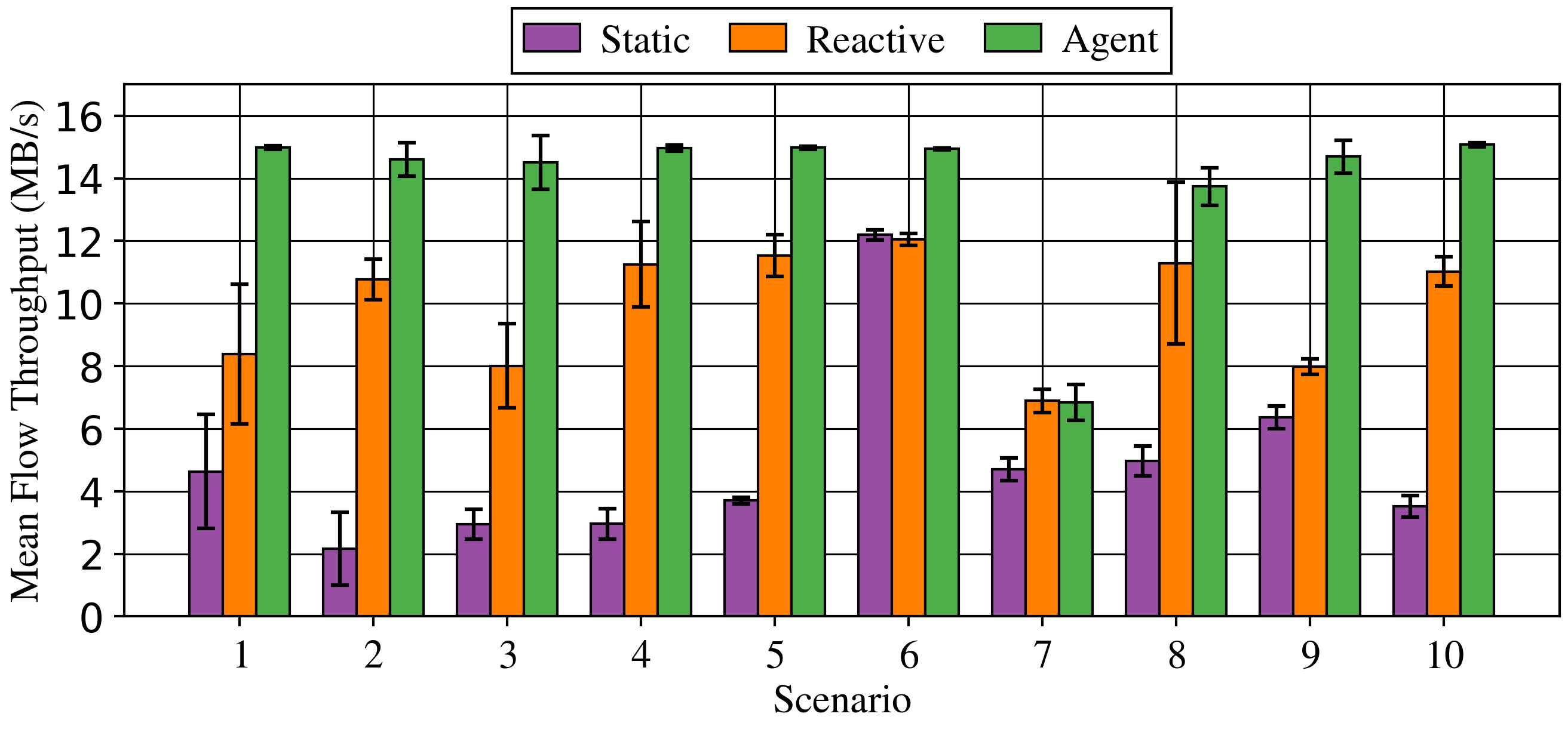}
\vspace{-0.1in}
\caption{Mean throughput of $f_p$ (MB/s) across the ten evaluation scenarios per policy (Static purple, Reactive orange, Agent green). Error bars are standard deviation across three repeats. The agent achieves the highest throughput in 9 of 10 scenarios, with dataset-wide means of 13.93 (agent), 9.92 (reactive), and 4.81 (static) MB/s.}
\label{fig:mean_throughput}
\vspace{-0.1in}
\end{figure}

Figure~\ref{fig:mean_throughput} reports the mean throughput of $f_p$ over the 120\,s congestion window for all ten scenarios, averaged across three repeats per policy. The agent achieves the highest mean throughput in 9 of 10 scenarios, with dataset-wide means of 13.93\,MB/s (agent), 9.92\,MB/s (reactive), and 4.81\,MB/s (static), a 40\% improvement over reactive (95\% CI [28\%, 56\%]) and a 3$\times$ improvement over static. Averaged across the ten scenarios, the agent's per-scenario advantage over reactive is $4.01$\,MB/s ($95\%$ CI $[2.48, 5.54]$\,MB/s, paired $t$-test $t = 5.94$, $p < 0.001$), consistent with the relative gain above. Per-scenario values appear in Appendix~\ref{app:stats}.

The agent's advantage holds across qualitatively different scenario types. In the full-overlap scenarios (S1-S4), where all five congesters share $k^*$ from the start, the agent leads reactive by 3.7-6.6\,MB/s. The margin narrows under partial overlap (S5, S7) or gradual load build-up (S8), where reactive has more opportunity to recover before collapse. In S6 the rolling pattern never accumulates $\xi_{\text{th}}$, so reactive collapses to static, while the agent still reroutes proactively and leads by 2.9\,MB/s. S9 yields the largest single-scenario improvement (+6.7\,MB/s) since the burst pattern blocks reactive from firing on the first congestion period, and by the time it fires on the second the episode is nearly over.
The agent loses only S7, by 0.06\,MB/s, a near-tie attributable to the destination-selection failure discussed above.

Static is competitive only in S6, where the rolling pattern leaves $k^*$ congested for only a fraction of the episode window; elsewhere it sustains severely degraded throughput for the full 120\,s. Rerouting is therefore necessary, and the choice of \emph{when} to reroute is what separates the agent from reactive.

\paragraph{Proactive Lead Time}
\label{para:lead_time}

\begin{figure}[t]
\centering
\includegraphics[width=0.85\columnwidth]{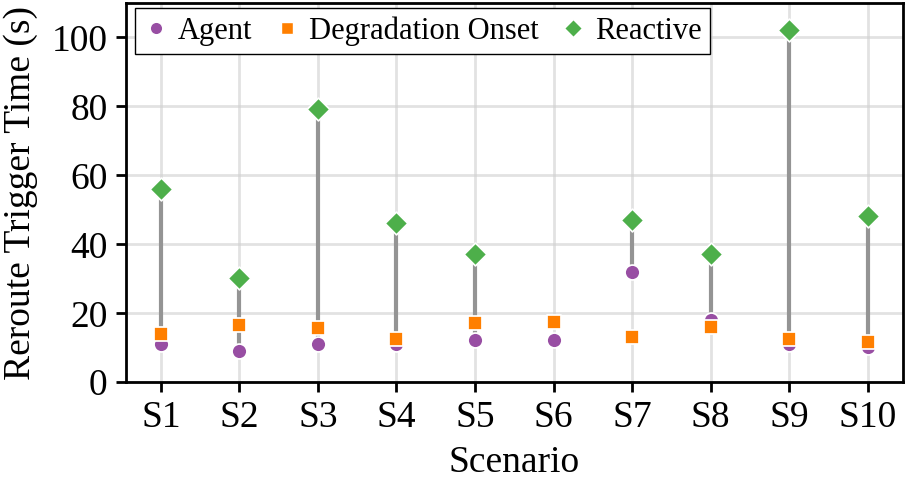}
\caption{Reroute trigger time (seconds after congestion start) for the agent and reactive policy across ten evaluation scenarios, shown as discrete per-scenario markers since the scenarios are independent. The shaded line for each scenario is the lead time gap between the agent and reactive. }
\label{fig:lead_time}
\vspace{-0.25in}
\end{figure}

Figure~\ref{fig:lead_time} reports the reroute trigger time for the agent and reactive policy across all ten scenarios, alongside a degradation-onset marker estimated from the steepest decline in the reactive policy's median throughput. The agent achieves a positive $\tau_{\text{lead}}$ in the majority of scenarios and repeats, rerouting while $f_p$ is still near its healthy baseline. Reactive achieves a negative $\tau_{\text{lead}}$ in every scenario where it fires, rerouting only after throughput has already collapsed.

The agent reroutes within 9-18\,s of congestion start across all scenarios, reactive fires 30-107\,s after start. The shaded bar at each scenario in Figure~\ref{fig:lead_time} represents the lead the agent gives over reactive in that scenario, with a mean of $34$\,s (95\% CI $[26.8, 41.5]$\,s, Appendix~\ref{app:stats}) across the eight scenarios where reactive fired on all three repeats. In S6 reactive never fires on any repeat (the rolling pattern clears each switch before $\xi_{k^*}$ can sustain above $\xi_{\text{th}}$ for the required consecutive window), and in S9 it fires on only two of three repeats, and only after 102-107\,s.

Agent reroute times are tight for most scenarios. In the steady-load scenarios (S1-S5, S10) all repeats fire within a 1-2\,s window with means in the 10-12\,s range. S8 is a mild outlier (mean 17\,s), where congesters arrive one at a time over 95\,s and the signal accumulates gradually rather than saturating. S7 varies more across repeats, two reroute at 9-11\,s and one at 32\,s, the repeat shown in Figure~\ref{fig:lead_time}, giving a scenario mean of 17\,s. Only two congesters share $k^*$ in this scenario, so the crowd signal sometimes builds more slowly. In both S7 and S8, the agent still leads reactive by over 30\,s on average, though $\tau_{\text{lead}}$ is reduced and is near-zero or negative in some individual repeats.

\paragraph{Token-Bucket Overflow at Reroute}
\label{para:tc_ol}

\begin{figure}[t]
\centering
\includegraphics[width=0.75\columnwidth]{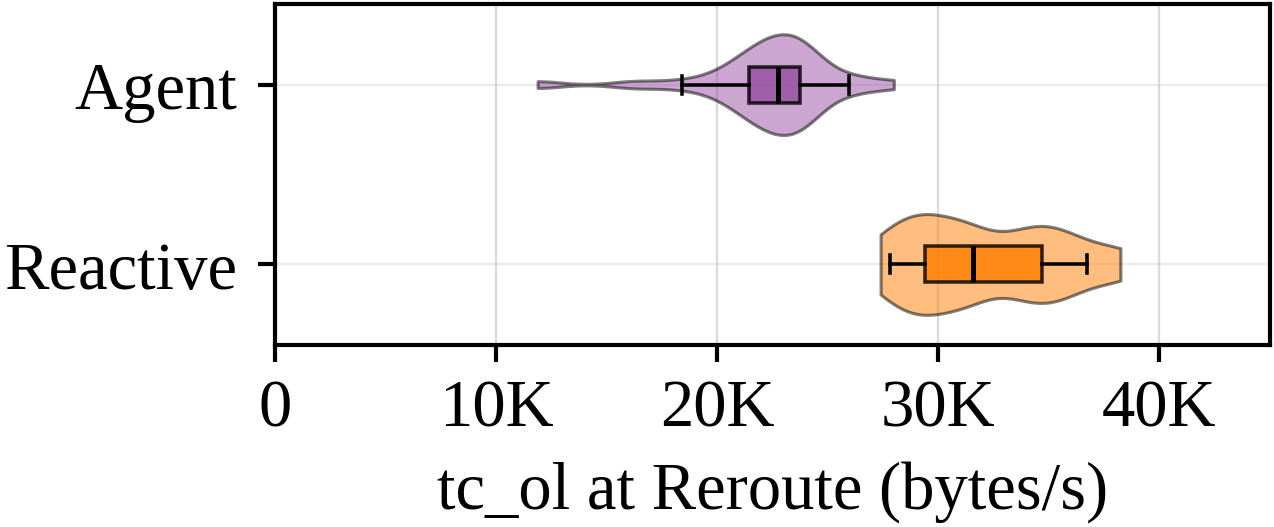}
\caption{Distribution of $\xi_{k^*}$ (bytes/s) at the moment of the first reroute, pooled across all scenarios and repeats. Agent has 30 data points, reactive has 26 (S6 never fired and one S9 repeat did not fire). Reactive fires by construction above $\xi_{\text{th}}$, while the agent fires well below.}
\label{fig:tcol_violin}
\vspace{-0.2in}
\end{figure}

Figure~\ref{fig:tcol_violin} shows the distribution of $\xi_{k^*}$ at the exact moment each policy issues its reroute, pooled across all scenarios and repeats. The two distributions are clearly separated. The agent fires at a mean $\xi_{k^*}$ of 22,300\,bytes/s, with most reroutes between 20,000-26,000\,bytes/s. Reactive fires at a mean of 32,100\,bytes/s, with all values above $\xi_{\text{th}} = 27{,}000$\,bytes/s by construction.
Reactive's mean exceeds $\xi_{\text{th}}$ because firing requires $\xi_{k^*} > \xi_{\text{th}}$ for three consecutive seconds (the smallest debounce that suppressed transient false positives in preliminary runs), during which $\xi_{k^*}$ continues to rise sharply from full-rate congester arrivals.

This separation corroborates the lead-time result. At the moment the agent reroutes, $\xi_{k^*}$ has not yet saturated, so $f_p$'s throughput is still near baseline. By the time reactive fires, the token bucket is in deep overflow and throughput has already collapsed. The agent's reroute times are driven by the crowd signal rather than by $\xi_{k^*}$. At the flow counts seen 10-17\,s after congestion start (typically 8-12 active flows on $k^*$), the crowd component of the learned value function is sufficient to trigger rerouting independently of $\xi_{k^*}$. The agent distribution has a low tail extending down to approximately 12,000\,bytes/s, corresponding to the earliest reroutes where the crowd signal crossed the decision boundary before $\xi_{k^*}$ had meaningful time to build. The one agent data point above $\xi_{\text{th}}$ corresponds to S7 repeat~1, where slower crowd-signal accumulation led to an unusually late reroute at 32\,s.

\paragraph{Q-Network Signal Analysis}
\label{para:probe}

To identify which signals drive the agent's rerouting decisions, we probe the locked Q-network by sweeping individual state dimensions through synthetic state vectors with all other dimensions held at baseline. Two sweeps are run, one varying the flow count on $k^*$ (crowd sweep) and one varying $\xi_{k^*}$ (overflow sweep). For each input we compute the decision margin $Q(\mathbf{s}, k^*) - \max_{k \neq k^*} Q(\mathbf{s}, k)$, where a positive margin means the agent prefers to stay and a negative margin means it prefers to reroute.

\begin{figure}[t]
\centering
\setlength{\tabcolsep}{1pt} 

\begin{tabular}{@{}cc@{}}
  \includegraphics[width=0.495\columnwidth]{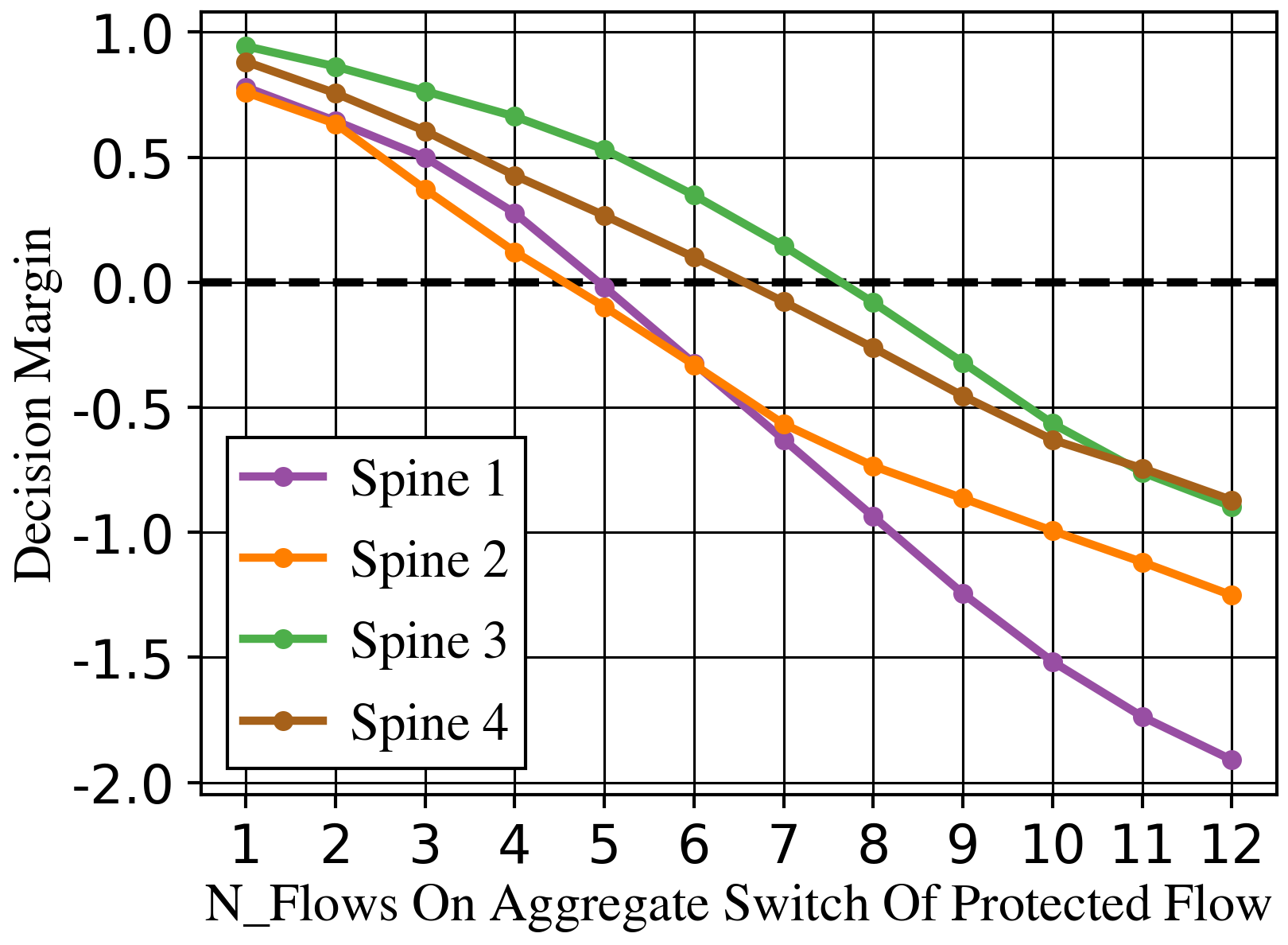} &
  \includegraphics[width=0.495\columnwidth]{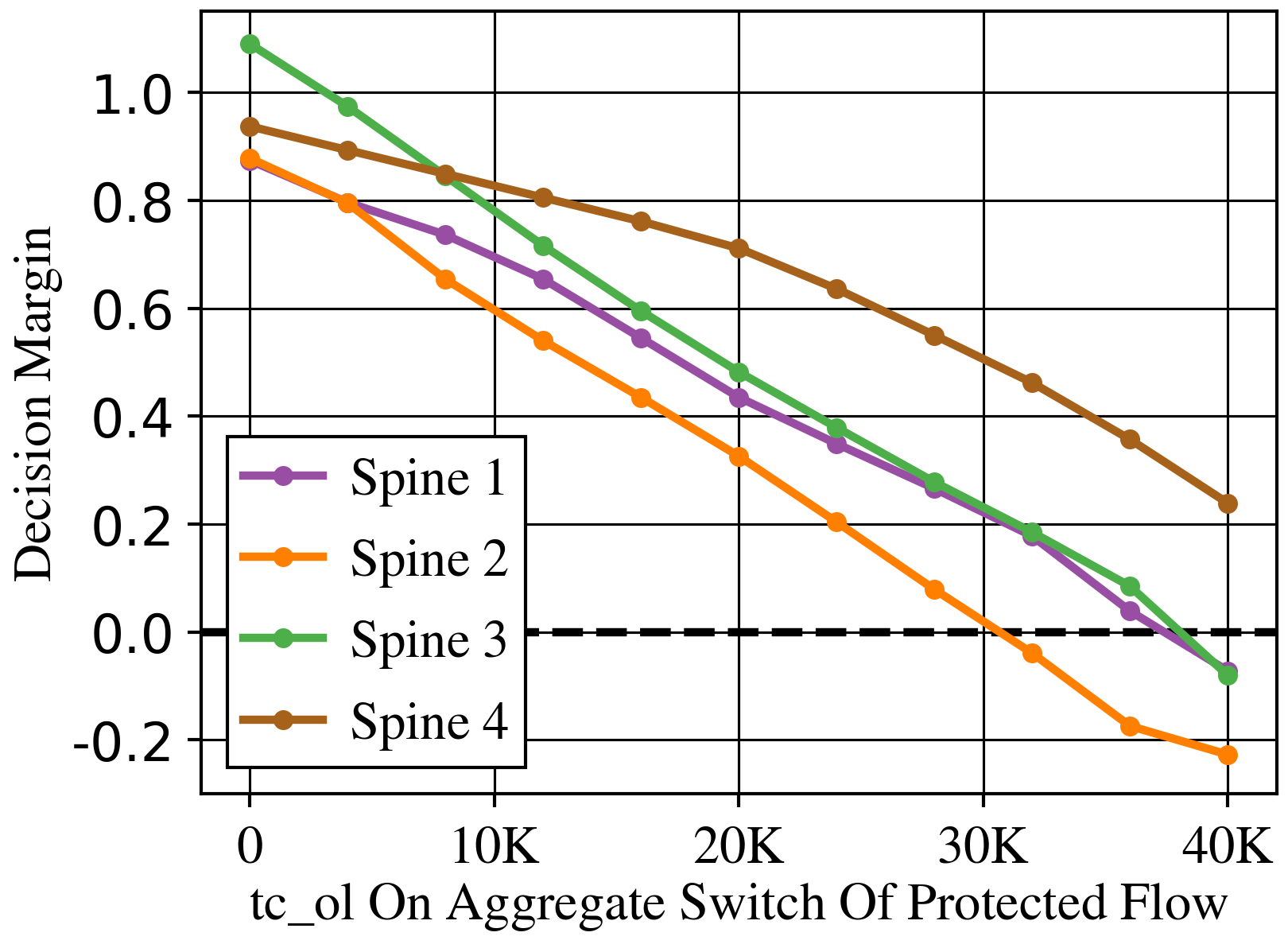} \\[-2pt]
  \small (a) Crowd Sweep &
  \small (b) \texttt{tc\_ol} Sweep
\end{tabular}
\vspace{-0.1in}
\caption{Q-network decision margin as a function of flow count with $\xi_{k^*}$ fixed at 5,000\,bytes/s (left) and as a function of $\xi_{k^*}$ with flow count fixed at 1 (right). A negative margin indicates a reroute decision. The crowd signal alone triggers rerouting at 5-8 flows depending on the switch, while $\xi_{k^*}$ alone does not trigger rerouting until 30,000-40,000\,bytes/s, above $\xi_{\text{th}}$ (dashed).}
\label{fig:probe}
\vspace{-0.2in}
\end{figure}

Figure~\ref{fig:probe} shows the results. In the crowd sweep (left panel), the margin crosses zero at a flow count of 5 for switches 1 and 2, and at 7-8 for switches 3 and 4, with $\xi_{k^*}$ held at a negligible 5,000\,bytes/s throughout. The crowd signal alone is therefore sufficient to flip the agent's decision at flow counts well below the maximum seen in live evaluation. In the overflow sweep (right panel), with the flow count fixed at 1, the margin does not cross zero until $\xi_{k^*}$ reaches 30,000-40,000\,bytes/s for switches 1-3, and never crosses zero for switch 4 within the swept range. Critically, at $\xi_{\text{th}}$ all four switch margins remain positive, so the agent would prefer to stay if $\xi_{k^*}$ were its only signal. Without the crowd signal, the agent would fire \emph{later} than reactive, not earlier. The probe provides a mechanistic explanation for the lead-time advantage. The agent acts on the crowd signal, which builds within seconds of congestion onset, rather than waiting for $\xi_{k^*}$ to accumulate to threshold-detectable levels.

\paragraph{Analysis of the Crowd Signal}
\label{para:crowd_ablation}

To quantify the empirical value of the crowd signal, we train a variant with the crowd penalty removed from the reward function and all other components identical. This ablated model can rely only on $\xi_{k^*}$ and throughput signals. Both models are evaluated on scenario C1, designed specifically to isolate the crowd signal.

C1 places 30 bandwidth-capped flows on $k^*$ (five congester nodes, six streams each, 1\,Mbps per stream) through the 200\,Mbps token bucket. The capped flows never saturate the bucket, so $\xi_{k^*}$ remains near the per-episode baseline and $\xi_{\text{th}}$ is never reached. The only signal that rises is the flow count, which jumps from 2 to over 30 within seconds of congestion start. C1 therefore acts as a binary test for crowd-signal learning, where a model that has learned the signal will reroute and one that has not will stay indefinitely.

\begin{figure}[t]
    \centering

    \includegraphics[width=0.78\columnwidth]{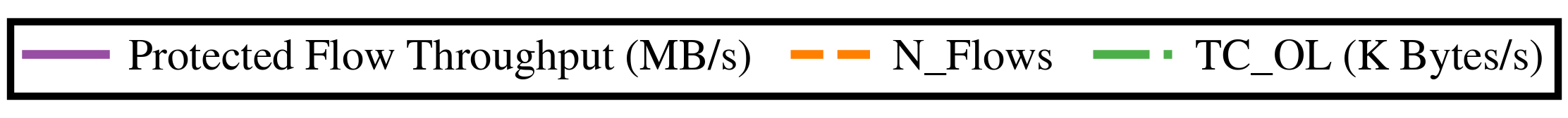}

    \vspace{-0.005em}

    \setlength{\tabcolsep}{0pt}
    \begin{tabular}{@{}c@{\hspace{0.005\columnwidth}}c@{}}
        \includegraphics[width=0.495\columnwidth]{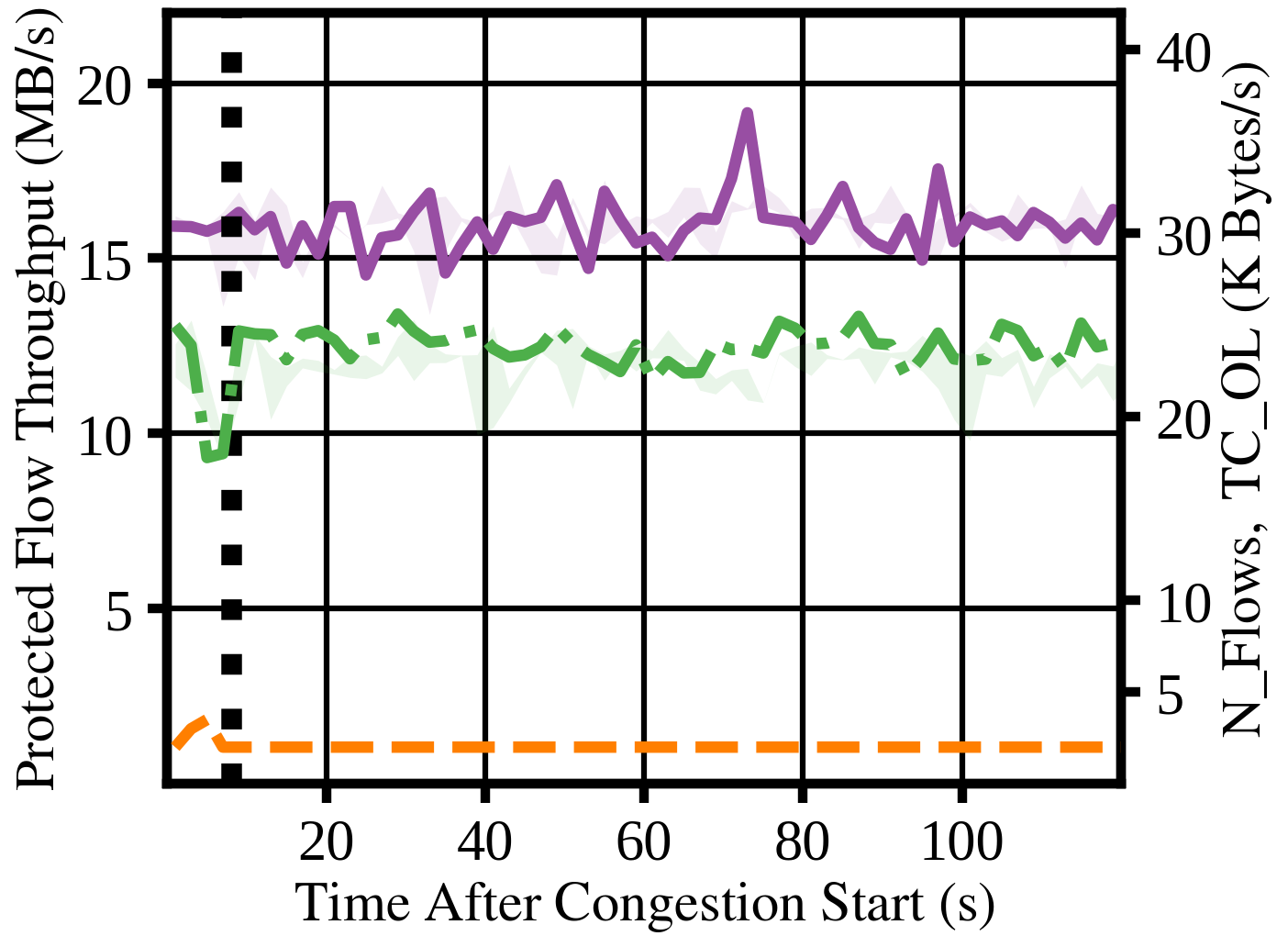} &
        \includegraphics[width=0.495\columnwidth]{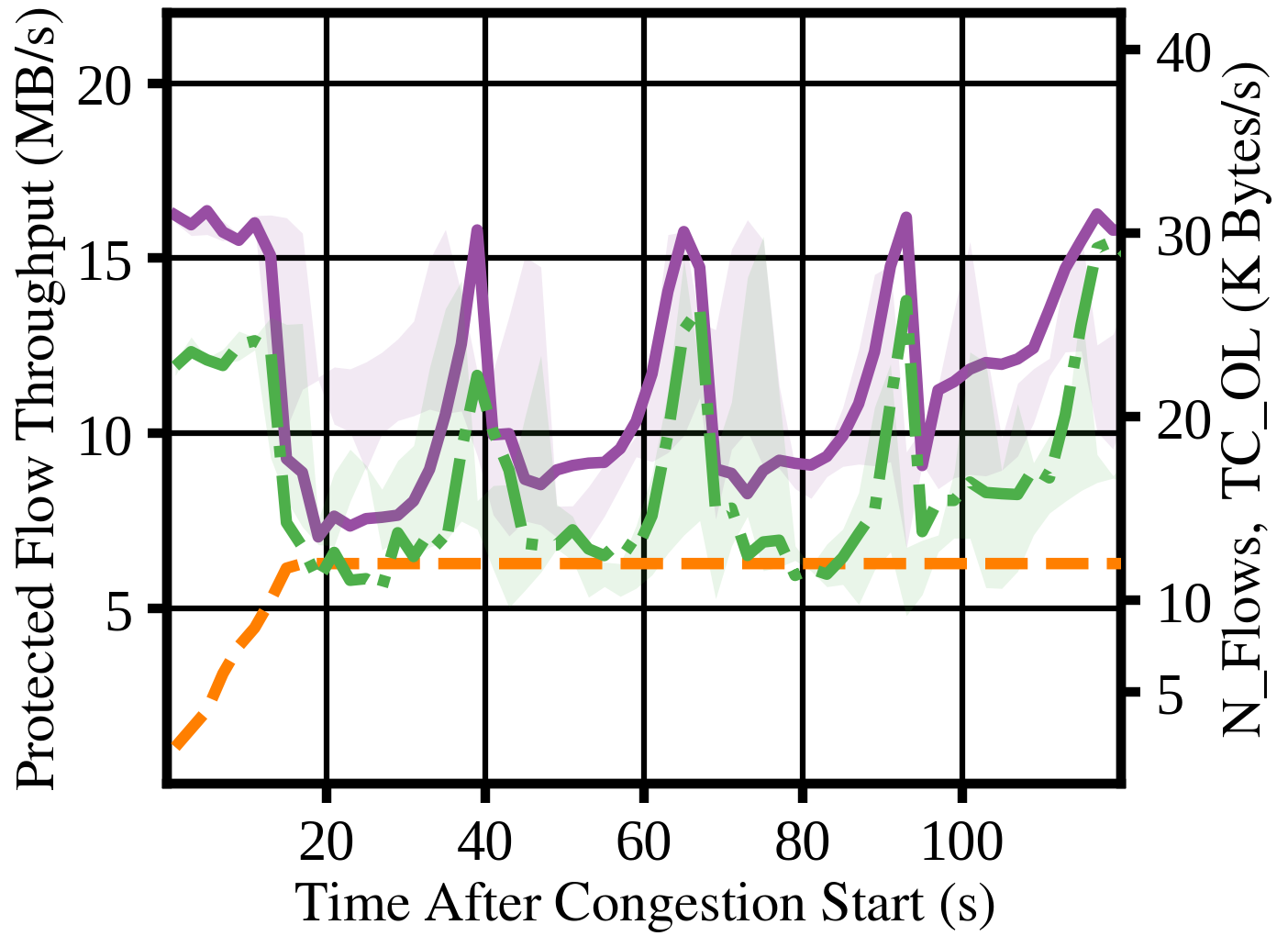} \\[-2pt]
        \small (a) With Crowd Signal &
        \small (b) Without Crowd Signal
    \end{tabular}

    \vspace{-0.4em}

    \caption{Throughput of $f_p$ on scenario C1 (stealth crowd, 30 bandwidth-capped flows, negligible $\xi_{k^*}$) for the full model (crowd signal active, left) and the ablated model (crowd signal removed, right). Median episode shown, with shaded bands giving the min-max range across the other two repeats. The vertical dotted line marks the reroute time in the full-model panel. The ablated model never reroutes.}
    \label{fig:ablation}
    \vspace{-0.2in}
\end{figure}

Figure~\ref{fig:ablation} shows the throughput timeseries for both models side by side. The full model detects the rising flow count, accumulates the required consecutive votes, and reroutes within approximately 12\,s while $f_p$ is still at near-baseline throughput. After rerouting, $\xi_{k^*}$ on the new switch stabilises at $f_p$'s own TCP baseline and throughput is maintained for the remainder of the episode. The ablated model has no mechanism to act on the flow count, never reroutes across any repeat, and lets $f_p$ compete with 30 flows for the full 120\,s, degrading and oscillating throughout. The mean throughput gap is 4.3\,MB/s (15.2 vs 10.9\,MB/s), the direct cost of removing the crowd signal in a stealth-crowd scenario where $\xi_{k^*}$ provides no usable information. On S1, where both $\xi_{k^*}$ and crowd signals are active, the gap narrows to 2.0\,MB/s (14.8 vs 12.8\,MB/s), consistent with the crowd signal providing a lead-time advantage on top of $\xi_{k^*}$ rather than being the sole trigger. While the crowd signal is necessary for the agent to detect stealth congestion, it is not sufficient on its own. Appendix \ref{app:crowd_scenarios} evaluates a \emph{flow-count threshold} as a standalone baseline and shows it fails in two special cases, one where it never fires and another where it selects the wrong destination.

\paragraph{Seed Robustness}
\label{para:seed_robustness}

\begin{figure}[t]
\centering
\includegraphics[width=0.7\columnwidth]{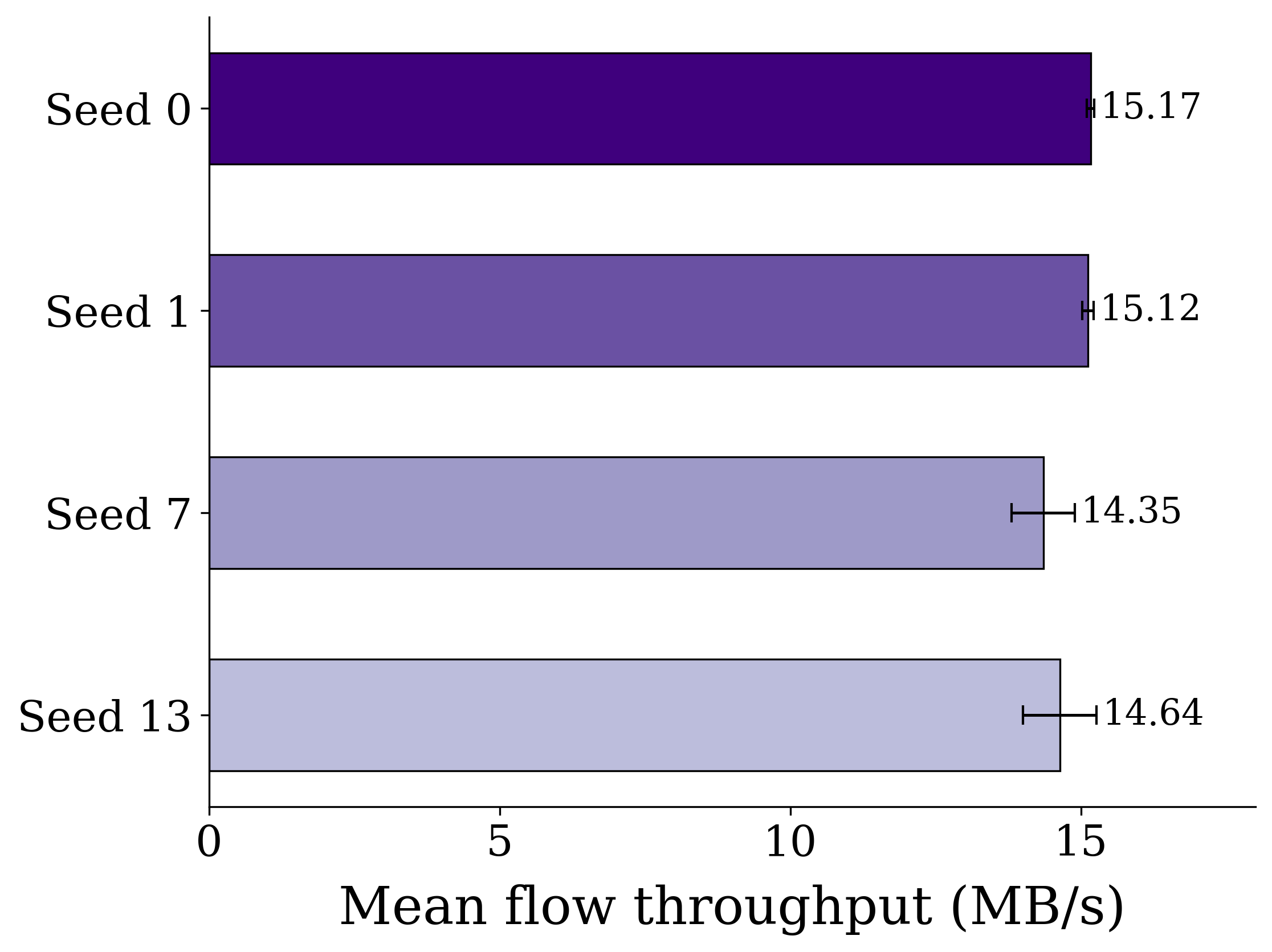}
\caption{Mean protected flow throughput on C1 for four independently trained model seeds. Error bars show standard deviation across three repeats. All seeds achieve 14.4--15.2\,MB/s, well above the A1\_no\_crowd baseline of 10.9\,MB/s (not shown), confirming that the crowd signal is stably learned across random initializations.}
\label{fig:seed_robustness}
\vspace{-0.2in}
\end{figure}

The crowd signal is a learned feature that emerges from the reward function and training data rather than being hardcoded. To verify that this emergence is not seed-specific, we train four additional models with different random seeds (identical hyperparameters, architecture, and corpus) and evaluate each on C1. Figure~\ref{fig:seed_robustness} shows the mean throughput for all four seeds. Every seed achieves between 14.4-15.2\,MB/s, well above the ablated-model baseline of 10.9\,MB/s. The spread across seed means is 0.8\,MB/s, comparable to within-seed run-to-run variability in the main evaluation. Since C1 is not in the training corpus and requires generalisation from full-rate congestion to a novel capped-flow pattern, the crowd signal is a stable, generalised feature of the learned policy rather than a seed-specific accident.

All datasets and related scripts along with a detailed README are provided as a GitHub repository \cite{google-drive}.

\section{Conclusions and Future Work}
\label{sec:conclusion}

This paper presented ProFlow, a proactive flow-placement framework for protecting performance-sensitive traffic in multi-tenant datacenter networks. By leveraging distributed telemetry signals and offline-trained reinforcement learning, ProFlow identifies precursor congestion conditions and reroutes a protected flow before throughput degradation occurs. The evaluation showed that ProFlow achieves approximately 40\% higher mean throughput than the reactive baseline while initiating rerouting decisions around 34 seconds earlier on average, demonstrating the effectiveness of anticipatory congestion management using early precursor signals. 

Future work will focus on extending ProFlow beyond single-flow protection toward intelligent multi-flow placement and protection. In practical datacenter environments, multiple performance-sensitive flows may simultaneously compete for limited clean paths, requiring the controller to jointly reason about congestion risk, path allocation, and flow prioritization across the network. This transforms the problem into a broader intelligent flow-placement and optimization problem under dynamic traffic conditions.


\bibliographystyle{ieeetr}
\bibliography{refs}

\appendices
\label{sec:appendix}
%

\appendices


\section{State Vector Construction}
\label{app:state}

This appendix gives the full per-coordinate construction of the state vector $\mathbf{s}_t \in \mathbb{R}^{d_s}$ summarised in Section~\ref{sec:design}, where $d_s = 6|\mathcal{S}| + 2|\mathcal{L}| + 4$ depends on the sizes of the aggregation-switch set $\mathcal{S}$ and the leaf-switch set $\mathcal{L}$ defined in Section~\ref{sec:background}. The constant $4$ is a fixed design dimension of $\mathbf{x}^{\text{flow}}_t$ that does not scale with the topology. Based on our implementation (Section~\ref{sec:implementation}), $|\mathcal{S}| = 4$ and $|\mathcal{L}| = 3$, giving the concrete state dimensions $\mathbf{x}^{\text{agg}}_t \in \mathbb{R}^{20}$, $\mathbf{x}^{\text{leaf}}_t \in \mathbb{R}^{6}$, $\mathbf{h}_t \in \{0,1\}^{4}$, $\mathbf{x}^{\text{flow}}_t \in \mathbb{R}^{4}$, and an overall state dimension of $d_s = 34$.

The first component, $\mathbf{x}^{\text{agg}}_t \in \mathbb{R}^{5|\mathcal{S}|}$, captures per-aggregation-switch congestion state across all aggregation switches:

\begin{equation}
\scriptsize
\mathbf{x}^{\text{agg}}_t =
\left[
\rho_k / C_\rho,\;
\mathrm{clip}(\dot{\rho}_k / C_\rho,-1,1),\;
\tilde{\xi}_k,\;
\tilde{n}_k,\;
e_k / C_\rho
\right]_{k \in \mathcal{S}}
\label{eq:agg_features}
\end{equation}

where $\rho_k$ denotes the transmit rate on switch $k$, $\dot{\rho}_k$ its rate of change, $\xi_k$ the token-bucket overflow rate (overlimits per second), $n_k$ the number of active flows, and $e_k$ the aggregate rate of elephant flows. The function $\mathrm{clip}(x,a,b)$ truncates its input to the interval $[a,b]$.

The normalised overflow and flow-count signals, $\tilde{\xi}_k$ and $\tilde{n}_k$, are computed differently depending on whether $k = k^*$. For the current host $k^*$, both signals are encoded as signed deviations from per-episode baselines:

\begin{equation}
\scriptsize
\tilde{\xi}_{k^*} = \mathrm{clip}\!\left(\frac{\xi_{k^*} - \bar{\xi}}{C_\xi},\, -1,\, 1\right), \qquad
\tilde{n}_{k^*}   = \mathrm{clip}\!\left(\frac{n_{k^*} - \bar{n}}{C_n},\, -0.5,\, 1\right)
\label{eq:signed_norm}
\end{equation}
where $\bar{\xi}$ is the per-episode mean token-bucket load contributed by the flow's own TCP traffic on its initial aggregation switch, and $\bar{n}$ is the per-episode baseline flow count on that aggregation switch before congestion arrives. Negative values of $\tilde{\xi}_{k^*}$ indicate that the flow's TCP output is collapsing below its own baseline, providing an early sign of severe congestion on the current path. Positive values of $\tilde{n}_{k^*}$ indicate that new flows have arrived above the per-episode baseline, which is the primary crowd signal used for proactive rerouting. For all other switches $k \neq k^*$, unsigned normalisations are used: $\tilde{\xi}_k = \xi_k / C_\xi$ and $\tilde{n}_k = n_k / C_n$, both clipped to $[0,1]$.

The second component, $\mathbf{x}^{\text{leaf}}_t \in \mathbb{R}^{2|\mathcal{L}|}$, captures precursor signals observed at the leaf switches in $\mathcal{L}$:

\begin{equation}
\scriptsize
\mathbf{x}^{\text{leaf}}_t =
\left[
\lambda_\ell / C_\lambda,\;
\mu_\ell / C_\rho
\right]_{\ell \in \mathcal{L}}
\label{eq:leaf_features}
\end{equation}

where $\lambda_\ell$ denotes the ingress rate of congestion flows at leaf $\ell$, and $\mu_\ell$ the corresponding transmit rate toward aggregation switches.

The third component, $\mathbf{h}_t \in \{0,1\}^{|\mathcal{S}|}$, is a one-hot encoding of the current placement $k^*$:

\begin{equation}
\scriptsize
(\mathbf{h}_t)_k = \mathbb{1}[\,k = k^*\,], \qquad k \in \mathcal{S}
\label{eq:placement_onehot}
\end{equation}

so that exactly one entry is set to one — corresponding to the current host $k^*$ — and all others are zero. This component allows the Q-network to condition on the current placement when interpreting the signed/unsigned signals in $\mathbf{x}^{\text{agg}}_t$.

The fourth component, $\mathbf{x}^{\text{flow}}_t \in \mathbb{R}^{4}$, captures flow-level intensity and flow health:

\begin{equation}
\scriptsize
\mathbf{x}^{\text{flow}}_t =
\left[
\phi_t / C_\phi,\;
F_t / C_\rho,\;
\mathrm{clip}(\Delta\phi_t / C_\phi,\,-1,\,1),\;
d_t
\right]
\label{eq:flow_features}
\end{equation}

where $\phi_t$ is the realised throughput of the protected flow on the current host (defined in Section~\ref{sec:background}), $F_t$ the aggregate rate of congestion flows, $\Delta\phi_t$ the change in $\phi_t$ since the previous step, and $d_t = \max(0,\,\bar{\phi}_t - \phi_t)\,/\,\bar{\phi}_t$ the fractional drop of $\phi_t$ relative to a rolling maximum baseline $\bar{\phi}_t$ computed over a short window.

Each scaling constant $C_x$ is a normalisation denominator chosen to map the typical operating range of the corresponding signal approximately to $[0,1]$, with values exceeding the constant clipped to $1$. Rate-based signals are normalised by the token-bucket rate limit of the aggregation switches, overflow indicators by an empirical upper bound observed across training scenarios, flow counts by a soft saturation threshold above which the crowd signal is treated as fully active, and throughput signals by the maximum observed flow rate of the protected flow.


\section{Synthetic Reward Function}
\label{app:reward}

This appendix gives the full form of the synthetic reward summarised in Section~\ref{sec:design}. The reward shares a common predictive base across both action types and adds action-gated shaping terms.

For a reroute action ($k \neq k^*$):

\begin{equation}
\scriptsize
\begin{aligned}
r^{\text{syn}}_t =\;&
\phi'_{\text{norm}}
- \mu_1\,\tilde{\xi}'_k
+ \mu_2\,\mathrm{ReLU}(\tilde{\xi}_{k^*} - \tilde{\xi}'_k) \\
&+ \mu_3\,(1 - \tilde{\xi}_{k^*})
- \mu_4\,\hat{\xi}^{\text{now}}_k
\end{aligned}
\label{eq:syn_reward_reroute}
\end{equation}

For a stay action ($k = k^*$):

\begin{equation}
\scriptsize
\begin{aligned}
r^{\text{syn}}_t =\;&
\phi'_{\text{norm}}
- \mu_1\,\tilde{\xi}'_{k^*}
+ \mu_2\,\mathrm{ReLU}(\tilde{\xi}_{k^*} - \tilde{\xi}'_{k^*}) \\
&+ \nu_1\,(1 - \tilde{\xi}_{k^*})
- \nu_2\,\tilde{\xi}_{k^*}
- \nu_3\,(1 - \phi'_{\text{norm}})
+ \beta\,\Delta n_{k^*}^{+}
\end{aligned}
\label{eq:syn_reward_stay}
\end{equation}

Primes denote values in the predicted next state, with $\phi'_{\text{norm}} = \phi'_{t+1}/\phi_0$ the predicted next-step throughput normalised by the per-episode pre-congestion baseline $\phi_0$ (cf.~Eq.~\ref{eq:real_reward}). The first three terms are common to both forms: $\mu_1\,\tilde{\xi}'_k$ penalises the predicted congestion of the chosen switch, and $\mu_2\,\mathrm{ReLU}(\tilde{\xi}_{k^*} - \tilde{\xi}'_k)$ is the proactive-escape bonus, which rewards moves toward a cleaner predicted state.

The reroute-specific terms in Eq.~\ref{eq:syn_reward_reroute} are the lead-time bonus $\mu_3\,(1 - \tilde{\xi}_{k^*})$, which fires while $k^*$ is still clean to incentivise early action, and the destination-congestion penalty $\mu_4\,\hat{\xi}^{\text{now}}_k$, which discourages rerouting to an already-congested switch.

The stay-specific terms in Eq.~\ref{eq:syn_reward_stay} are: $\nu_1\,(1 - \tilde{\xi}_{k^*})$ rewards remaining on a clean current switch; $\nu_2\,\tilde{\xi}_{k^*}$ penalises staying as the current switch becomes congested; $\nu_3\,(1 - \phi'_{\text{norm}})$ penalises predicted throughput collapse; and $\beta\,\Delta n_{k^*}^{+}$ ($\beta < 0$) is the crowd penalty consistent with the real-transition reward (Eq.~\ref{eq:real_reward}).


\section{Training Strategies}
\label{app:strategies}

Table~\ref{tab:training_strategies} lists the full set of training-data strategies summarised in Section~\ref{sec:implementation}. Each strategy specifies (i) where the protected flow $f_p$ is initially placed, (ii) where the congester nodes generate traffic, and (iii) whether and when the data-collection runner issues a reroute. Together the strategies span the full reward-design space of Section~\ref{sec:design}: the base strategies (A--E) cover escape from a congested switch, placement diversity, stay-on-clean behaviour, the precursor window required to learn proactive action, and full pile-up; the dwell-time variants (A\_LONG, A\_SHORT, D\_SHORT, D\_LONG) sweep reroute timing within those base setups; and the late-addition strategies (F\_STAY, F\_LATE, F\_EARLY, G\_CLEAN, H\_PARTIAL) target stay-vs-leave decisions, positional bias, and tolerance of partial congestion that earlier datasets under-represented. The 779-episode corpus referenced in Section~\ref{sec:implementation} is built from these strategies across dataset versions v4, v6, v7, v8, and v10.

\begin{table}[t]
\centering
\caption{Training episode strategies. Each strategy defines how congestion is placed relative to the protected flow and whether or when the data-collection runner issues a reroute. Strategies A--E were collected in datasets v4--v7; F--H in v8; A\_SHORT and F\_EARLY in v10.}
\label{tab:training_strategies}

\footnotesize
\renewcommand{\arraystretch}{1.15}

\begin{tabularx}{\columnwidth}{lX}
\toprule
\textbf{Strategy} & \textbf{Description} \\
\midrule

A &
Protected flow on the most-loaded switch; runner reroutes once tc\_ol is confirmed above threshold. Teaches escape from a congested switch. \\

A\_LONG &
Same as A but with an extended dwell (40--60\,s) before rerouting. \\

A\_SHORT &
Same as A but reroute issued early (5--20\,s), balanced across all four switches. Restores correct per-switch departure timing. \\

B &
Protected flow on a randomly selected switch; reroute at a random time. Provides placement and timing diversity. \\

C &
Protected flow on the cleanest switch; no reroute. Teaches that staying on a clean switch yields high reward. \\

D &
Protected flow starts on a clean switch; 2--3 congesters are subsequently rerouted onto it. Runner waits for tc\_ol to rise before rerouting. Creates the pre-congestion precursor window that is the primary training signal for proactive behaviour. \\

D\_SHORT &
Same as D with a shorter post-arrival dwell (10--20\,s). \\

D\_LONG &
Same as D with a longer post-arrival dwell (30--50\,s). \\

E &
All 5 congesters on the protected flow's switch from the start; full pile-up. Runner reroutes after 20--40\,s. Teaches response at the high end of the tc\_ol range. \\

F\_STAY &
All 5 congesters on the protected flow's switch; no reroute issued. Teaches that high tc\_ol does not always mandate rerouting. \\

F\_LATE &
All 5 congesters on the protected flow's switch; reroute issued late (40--60\,s). \\

F\_EARLY &
All 5 congesters on the protected flow's switch; reroute issued early (5--20\,s), balanced across all four switches. Teaches early departure under maximum load. \\

G\_CLEAN &
All 5 congesters on a \emph{different} switch from the protected flow; no reroute. Corrects positional bias by teaching that any clean switch warrants staying. \\

H\_PARTIAL &
1--3 congesters on the protected flow's switch; no reroute. Teaches tolerance of partial congestion without unnecessary rerouting. \\

\bottomrule
\end{tabularx}
\end{table}


\section{Evaluation Scenarios}
\label{app:scenarios}

Table~\ref{tab:scenarios} lists the full set of evaluation scenarios summarised in Section~\ref{sec:implementation}. The ten main scenarios (S1--S10) are designed to vary independently along three axes: \emph{spatial overlap} between the protected flow $f_p$ and the congesters (full overlap S1--S4, partial overlap S5/S7, no overlap by construction in C1/C2), \emph{temporal profile} of the congestion (static S1--S4, rolling S6, ramp-up S8, burst S9, heavy S10), and \emph{topological position} of the protected flow (S1--S4 cycle through all four aggregation switches in $\mathcal{S}$). The two ablation scenarios (C1, C2) isolate the crowd-driven component of the agent's behaviour by capping per-flow bandwidth, so that token-bucket overflow remains negligible and only $\Delta n_{k^*}^{+}$ can drive a reroute.

\begin{table}[t]
\centering
\caption{Evaluation scenarios. All main scenarios (S1--S10) use 5 congester nodes running full-rate TCP flows unless stated otherwise. Each scenario is evaluated under 3 policies (agent, reactive, static) $\times$ 3 repeats = 9 episodes per scenario. Ablation scenarios (C1, C2) use 3 repeats per model variant. All episodes run for 120\,s post-congestion start.}
\label{tab:scenarios}

\renewcommand{\arraystretch}{1.15}

\begin{tabularx}{\columnwidth}{lX}
\toprule
\textbf{Scenario} & \textbf{Description} \\
\midrule
S1  & All 5 congesters on the same aggregation switch as the protected flow (Switch 1), full-rate \\
S2  & Same as S1 but protected flow and all congesters on Switch 2 \\
S3  & Same as S1 but protected flow and all congesters on Switch 3 \\
S4  & Same as S1 but protected flow and all congesters on Switch 4 \\
S5  & Partial overlap: 3 of 5 congesters share the protected flow's switch; 2 are on other switches \\
S6  & Rolling: congesters start on the protected flow's switch and migrate to a new switch every 30\,s \\
S7  & Minority overlap: only 2 of 5 congesters share the protected flow's switch; 3 are elsewhere \\
S8  & Ramp-up: all congesters target the protected flow's switch but arrive one at a time every 25\,s \\
S9  & Burst: congesters arrive simultaneously, clear after 30\,s, and return after another 30\,s \\
S10 & Heavy load: same placement as S1 but each congester runs double the number of streams \\
\midrule
C1  & Stealth crowd: 30 bandwidth-capped flows on the protected flow's switch --- n\_flows rises sharply but tc\_ol remains near zero \\
C2  & Slow-ramp crowd: bandwidth-capped flows arrive one at a time every 10\,s --- isolates adaptive baseline behaviour \\
\bottomrule
\end{tabularx}

\end{table}



\section{Reactive Threshold Selection}
\label{app:threshold}

The reactive baseline reroutes the protected flow when the token-bucket overflow rate $\xi_{k^*}$ on its current aggregation switch $k^*$ stays above a fixed threshold $\xi_{\text{th}}$ for three consecutive samples. The protected flow runs at about $92\%$ of its token-bucket rate, so it produces some overflow even when no congesting flows are present. The threshold must sit above this self-induced overflow, or the policy false-alarms on the flow's own traffic. This appendix shows that $\xi_{\text{th}} = 27{,}000$ bytes/s is the lowest threshold that never triggers on this self-induced overflow.

We use $120$ episodes collected on the original testbed. A sample counts as self-induced overflow when the protected flow is already at full rate on its current aggregation switch $k^*$ and no congesting host shares that switch ($n_{k^*} \leq 2$, the protected flow alone). We only count a sample once the flow's rate is above $15\times10^6$ bytes/s, since at the start of each episode the connection is still ramping up to full speed and has not yet reached this steady state. Any reactive firing under these conditions is a false alarm, since no congestion is present. This gives $5{,}433$ samples across $107$ episodes. The remaining $13$ episodes never place the protected flow alone on its switch at full rate, so they contribute no self-noise samples.

\begin{figure}[tb]
  \centering
  \begin{tabular}{@{}cc@{}}
    \includegraphics[width=0.495\columnwidth]{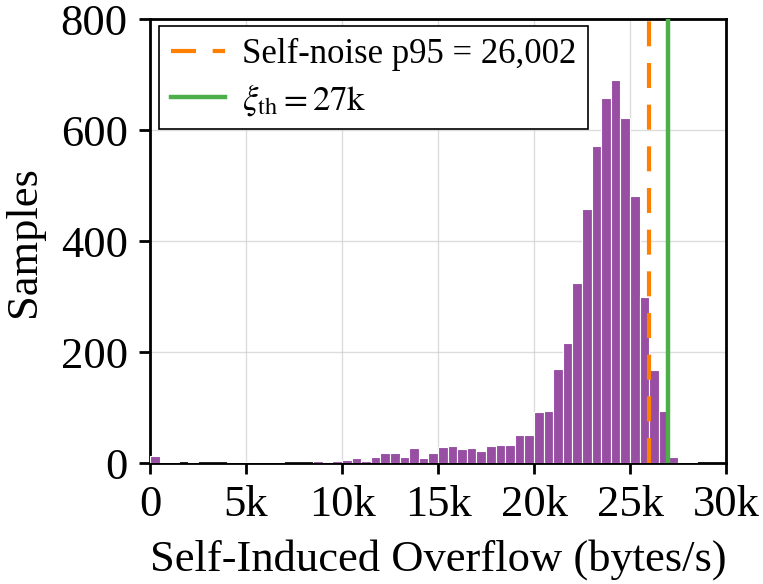} &
    \includegraphics[width=0.495\columnwidth]{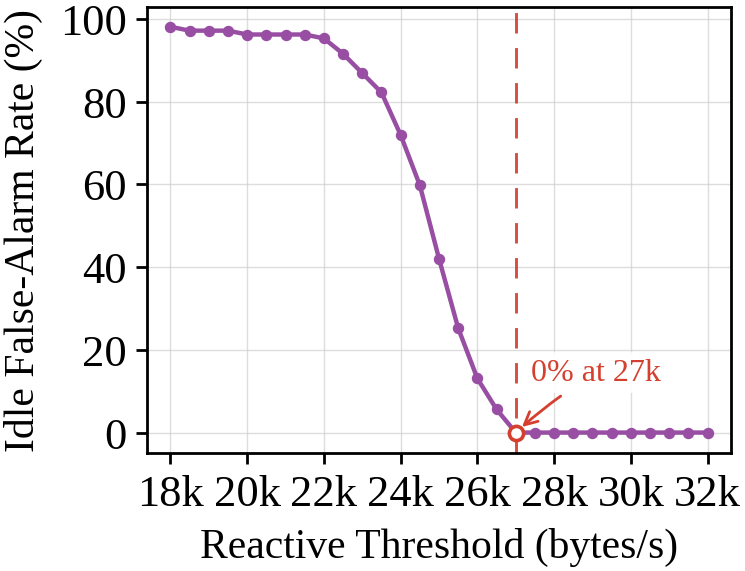} \\[-2pt]
    \small (a) Self-induced overflow & \small (b) False-alarm rate
  \end{tabular}
  \caption{(a) Distribution of the protected flow's self-induced token-bucket overflow ($5{,}433$ samples across $107$ episodes), with the $95$th percentile (dashed) and the chosen threshold $\xi_{\text{th}} = 27{,}000$ bytes/s (solid) marked. (b) False-alarm rate of the reactive policy against the overflow threshold in the no-congestion regime, reaching $0\%$ at $27{,}000$ bytes/s.}
  \label{fig:threshold-selfnoise}
\end{figure}

Figure \ref{fig:threshold-selfnoise}(a) shows the distribution of this self-induced overflow. It concentrates between $22{,}000$ and $26{,}000$ bytes/s, with a mean of $23{,}093$ bytes/s, matching the steady-state baseline reported in Section \ref{sec:implementation}. The distribution has a heavy upper tail, with a $95$th percentile of $26{,}002$ bytes/s and occasional spikes up to $28{,}336$ bytes/s. A threshold placed inside this tail would be tripped by the protected flow alone.

For each candidate threshold, we apply the reactive rule exactly, firing whenever the overflow exceeds the threshold for three consecutive samples, and count how many of the $107$ episodes would false-alarm in the no-congestion case above. Figure~\ref{fig:threshold-selfnoise}(b) plots this false-alarm rate. It falls from $42.1\%$ at $25{,}000$ bytes/s to $13.1\%$ at $26{,}000$ bytes/s and reaches $0\%$ at $27{,}000$ bytes/s, the lowest threshold with zero false alarms at $500$ byte resolution. The chosen value $\xi_{\text{th}} = 27{,}000$ bytes/s is therefore the tightest setting that clears the protected flow's own overflow while staying as sensitive as possible to real congestion. A lower threshold causes spurious reroutes, and a higher one only delays the reactive response.




\section{Crowd Baseline on Rate-Count Decoupling}
\label{app:crowd_scenarios}

\begin{table}[t]
\centering
\caption{Rate-count-decoupling scenarios evaluated against the crowd baseline. S11 and S12 add a fourth policy, crowd, which reroutes when the flow count on the protected flow's switch exceeds a fixed threshold $T=5$, and are constructed to test where this flow-count signal diverges from the true congestion state. Each scenario is evaluated under 3 repeats per policy.}
\label{tab:crowd_scenarios}
\renewcommand{\arraystretch}{1.15}
\begin{tabularx}{\columnwidth}{lX}
\toprule
\textbf{Scenario} & \textbf{Description} \\
\midrule
S11 & Elephant: a single congester saturates the protected flow's switch at full rate; n\_flows reaches only 4, below the flow-count threshold, while tc\_ol and loss rise normally \\
S12 & Destination trap: the protected flow's switch is congested and every candidate switch reports the same n\_flows, but one of them is saturated by a single elephant flow \\
\bottomrule
\end{tabularx}
\end{table}

The flow count $n_k$ counts distinct source-destination pairs on switch $k$ and therefore scales with the number of congesters present rather than their transmission rate. Scenarios S1--S10 hold the congester count fixed at five, so flow count and offered load increase together and either signal alone is sufficient to detect congestion. Scenarios S11 and S12, summarised in Table~\ref{tab:crowd_scenarios}, decouple these two quantities. In S11 a single congester saturates $k^*$ at the offered load level of S1, while $n_{k^*}$ reaches only four. In S12 every candidate switch reports an equal flow count while one of them is saturated, so $n_k$ provides no information for destination selection. Both scenarios are deliberate stress tests of the flow-count signal rather than representative workloads and are reported separately from S1--S10 for that reason. The crowd baseline is evaluated only on S11 and S12. S11 and S12 were also collected on a different FABRIC site than S1--S10, a consequence of FABRIC's limited lease durations and the resulting unavailability of the original site rather than an experimental choice. Both sites realise the same logical topology described in Section~\ref{sec:implementation}, and the agent evaluated on S11 and S12 is the same model trained on the original deployment, with no additional training performed on the new site. This successful rerouting on the new site suggests \emph{that the learned policy is not specific to the hardware of the original deployment and generalises across physical sites}.

\begin{figure}[t]
\centering
\begin{tabular}{cc}
  \includegraphics[width=0.47\columnwidth]{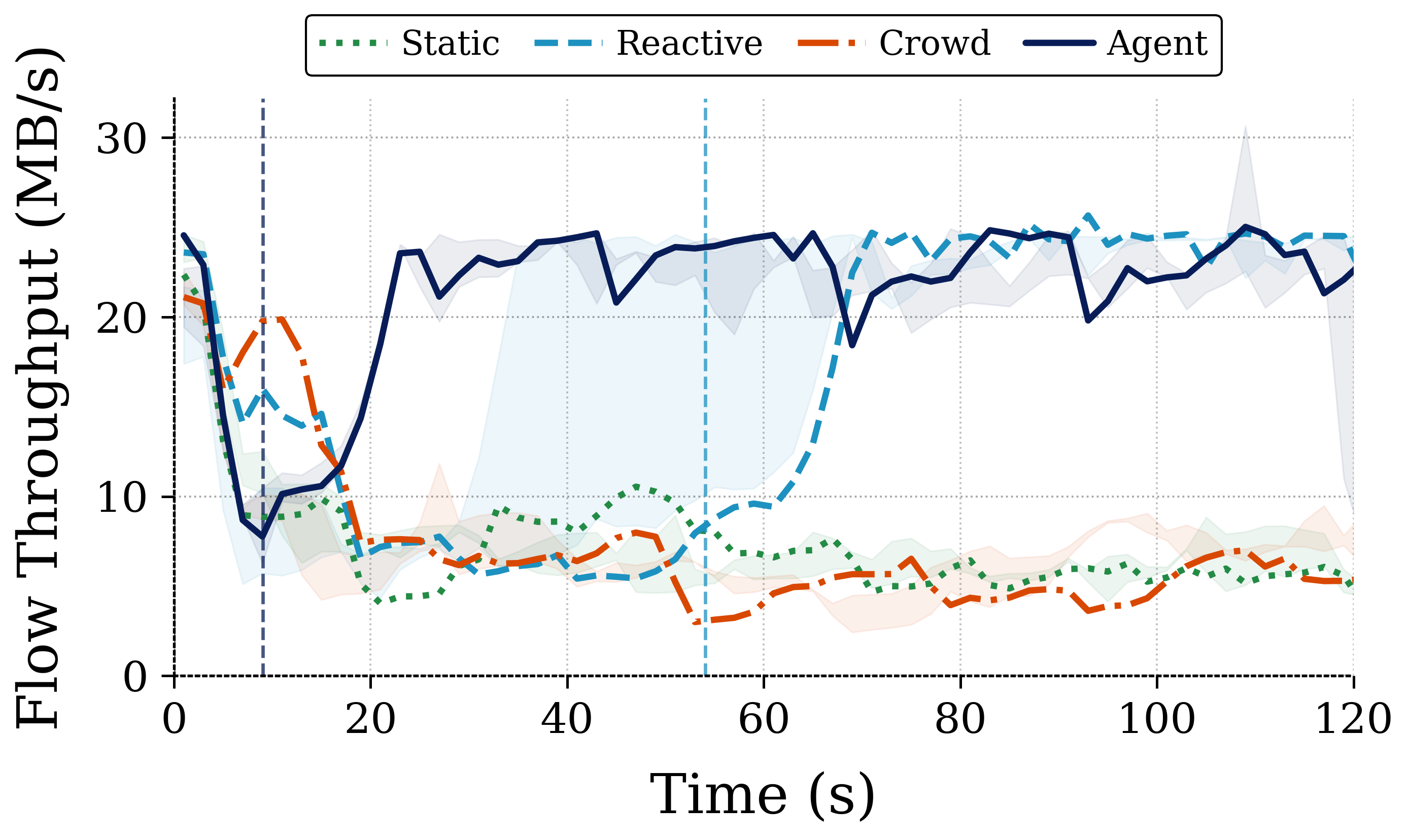} &
  \includegraphics[width=0.47\columnwidth]{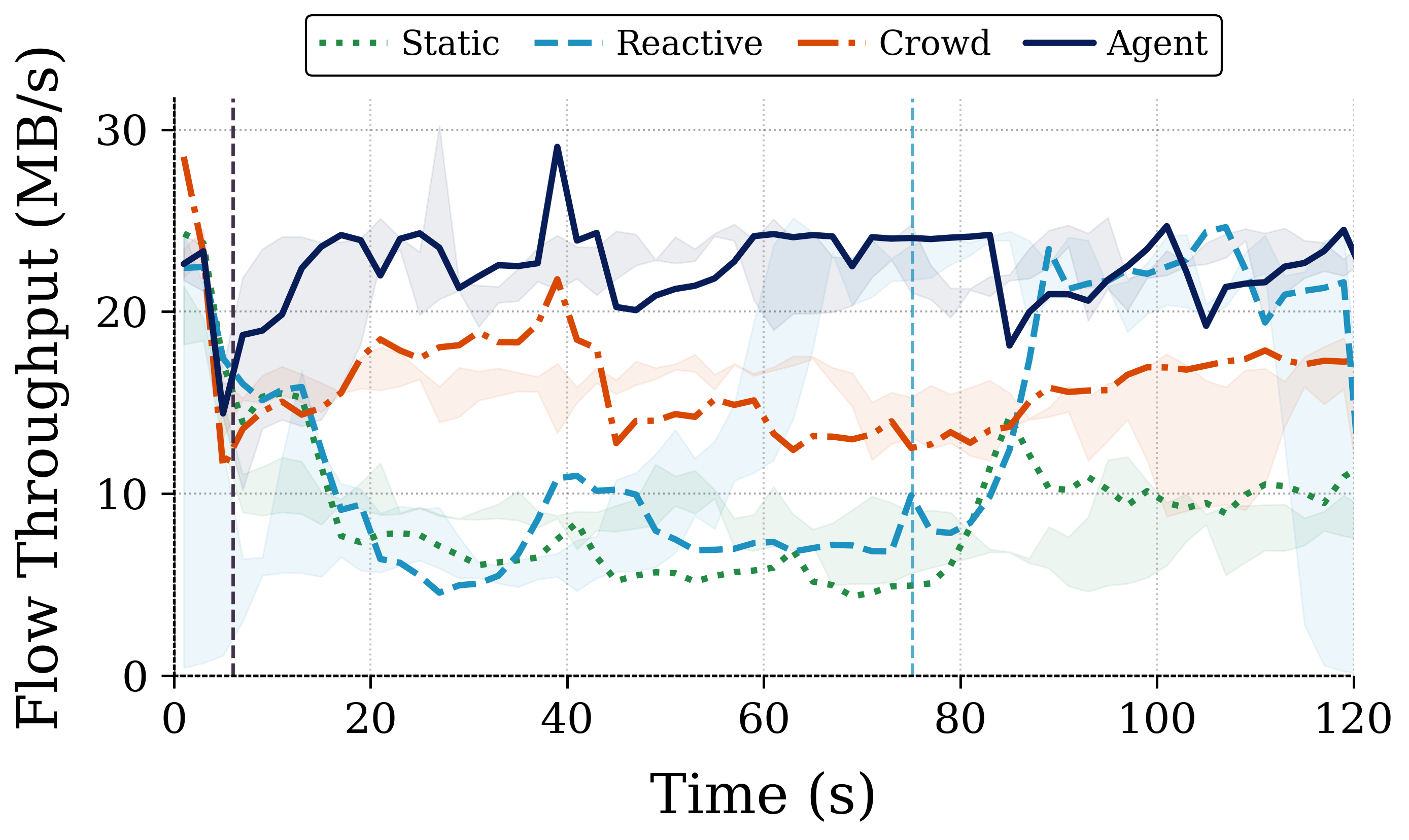} \\[-2pt]
  \small (a) S11: Elephant & \small (b) S12: Destination Trap \\
\end{tabular}
\caption{Throughput of $f_p$ (MB/s) over the congestion window for S11 (a) and S12 (b), with the Crowd baseline added (Static: green dotted; Reactive: blue dashed; Crowd: orange dash-dot; Agent: navy solid). Each panel shows the median episode per policy, with shaded bands giving the min-max range across the other two repeats and a vertical dashed marker at the first reroute of the median episode.}
\label{fig:timeline_s11_s12}
\end{figure}

Figure~\ref{fig:timeline_s11_s12} shows the throughput of $f_p$ over the congestion window for S11 and S12 under all four policies. In S11 the single elephant congester drives the flow count on $k^*$ to only four, below the crowd threshold $T=5$, so Crowd never fires and tracks Static near the noise floor, while the agent reroutes at approximately nine seconds and holds $f_p$ near its healthy baseline. In S12 Crowd and the agent both reroute at approximately six seconds, but Crowd selects the elephant-loaded destination switch because it reports the minimum flow count and recovers only partially, whereas the agent reads the elevated overflow $\tilde{\xi}_k$ on that switch and routes to a clean destination instead. Reactive recovers in both scenarios, but only after $\xi_{k^*}$ sustains above $\xi_{\text{th}}$ for three seconds, well after the agent has already recovered. Together these two timelines show that a flow-count threshold alone is insufficient for proactive rerouting, since it can fail to trigger when congestion arises from a single high-rate flow and can trigger correctly but still select the wrong destination when flow count does not distinguish between candidate switches.

\section{Crowd Threshold Selection}
\label{app:crowd}

This appendix justifies the crowd signal's flow-count threshold $T = 5$. The crowd signal reroutes the protected flow when the flow count $n_{k^*}$ on its current aggregation switch exceeds a threshold $T$. We test this threshold on three scenarios, S1, S5 and S8, each stressing the flow count in a different way. In S1, all five congesting hosts join at once, so the flow count rises quickly and clears any reasonable threshold. This is the control case, where the threshold never matters. In S5, only three of the five congesters share the switch, so the flow count stays close to the threshold, making this the case most sensitive to $T$. In S8, congesters arrive one at a time, so the flow count rises slowly, and exactly when it crosses $T$ depends on where $T$ is set. Together, these three cases cover a high, a marginal, and a slowly rising flow count, the three situations where a count threshold could fail.

\begin{figure}[tb]
  \centering
  \begin{tabular}{@{}cc@{}}
    \includegraphics[width=0.495\columnwidth]{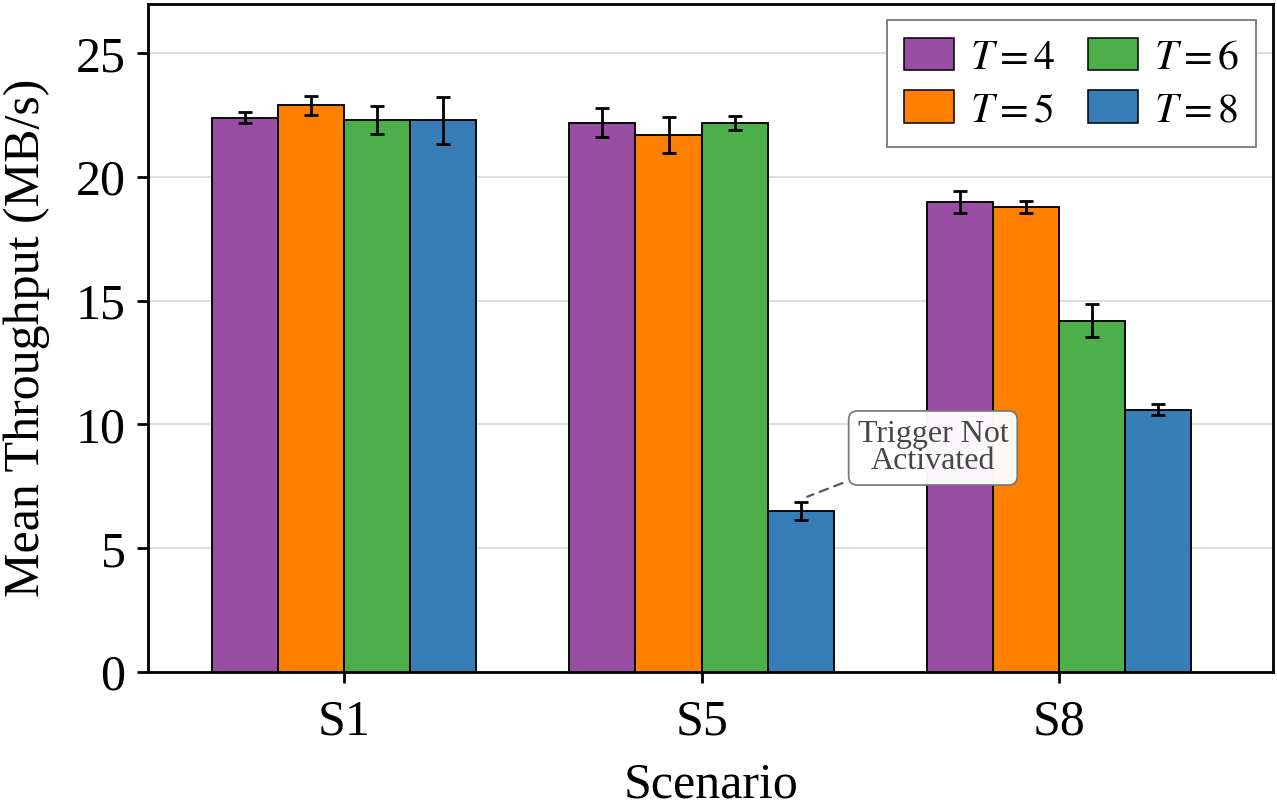} &
    \includegraphics[width=0.495\columnwidth]{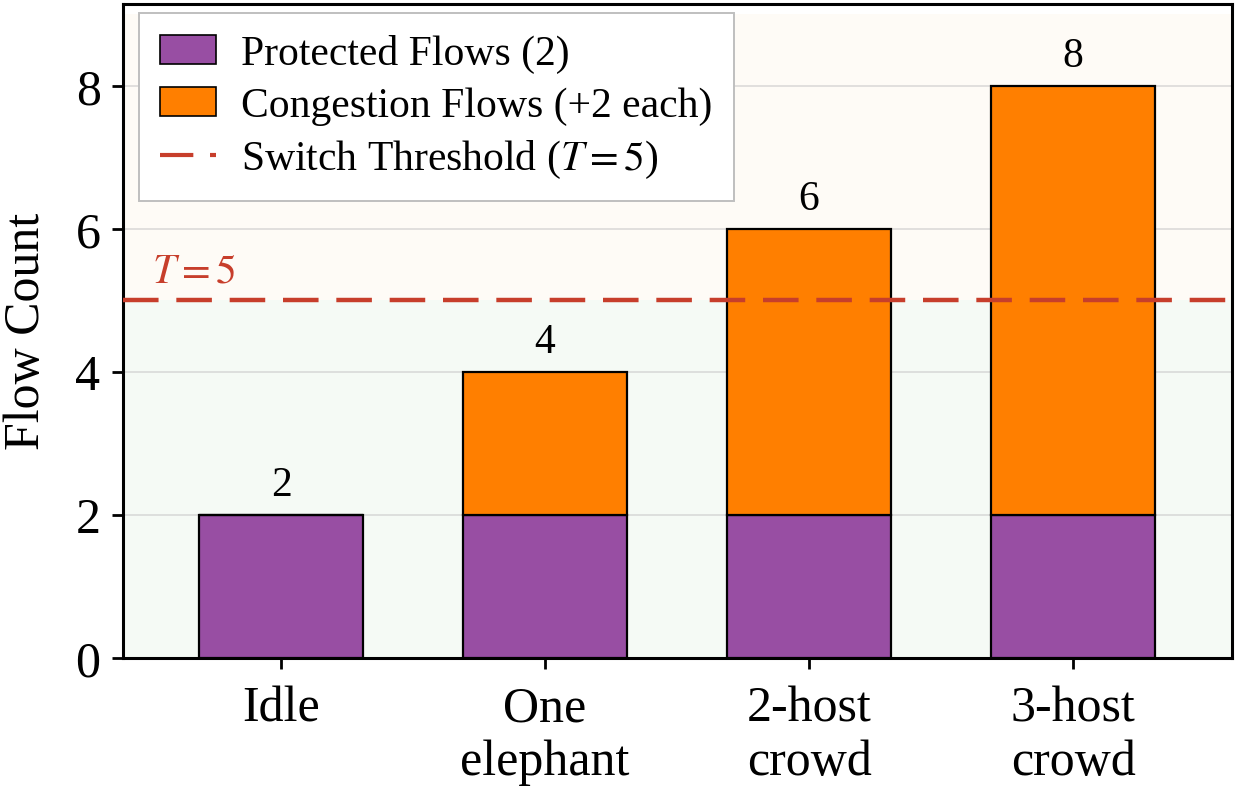} \\[-2pt]
    \small (a) Count Threshold sweep & \small (b) Count Structure
  \end{tabular}
  \caption{(a) Crowd-signal mean throughput as the flow-count threshold $T$ changes, for a high (S1), marginal (S5), and slowly rising (S8) flow count. Three repeats per bar, error bars show one standard deviation. (b) The flow count $n_k$ on the protected flow's switch adds up, two entries for the protected flow itself plus two for every congesting host present.  }
  \label{fig:crowd-tuning}
\end{figure}

Figure~\ref{fig:crowd-tuning}(a) shows the crowd policy's mean throughput as $T$ changes. In S1, throughput stays flat at about $22$ MB/s for every $T$, since the flow count is already high enough that the threshold never gets in the way. In S5, throughput holds steady until $T = 8$, where it drops sharply to $6.5$ MB/s. In S8, throughput falls more gradually, from $19$ MB/s at $T \in \{4, 5\}$ down to $10.6$ MB/s at $T = 8$. In both S5 and S8, raising $T$ only makes the crowd signal perform worse.

The flow count $n_k$ that this threshold reads is not continuous, and it does not depend on how much traffic a flow sends. Each connection installs two entries in the switch's flow table, one for each direction. So the protected flow alone always contributes two entries, and each congesting host adds two more. This means $n_k$ can only take the values $4$ with one congesting host, $6$ with two, and $8$ with three. Because $n_k$ does not depend on rate, a single congesting flow always produces $n_k = 4$, whether that flow is harmless or, as in scenario S11 (Appendix~\ref{app:crowd_scenarios}), a real elephant flow causing genuine congestion. Either way, it is still one flow, not a crowd. Elephant runs (S11) always peak at $n_k = 4$, crowd runs (S5 and S8) always reach $n_k = 8$ or $12$, and no run ever produces $n_k = 5$. Since a single flow can produce $n_k = 4$, the threshold must not fire at that count, which sets a lower bound of $T \geq 4$. We set $T = 5$ rather than $T = 4$ for the safety margin this adds; $T = 4$ would sit exactly on the single-flow count with no room to spare, while $T = 5$ places the threshold a full step above it, in the gap between $n_k = 4$, a single flow, and $n_k = 6$, the smallest real crowd, so it fires on every real crowd but never on a single flow, as Figure~\ref{fig:crowd-tuning}(b) shows. This margin costs nothing, across all three scenarios the mean throughput at the two thresholds differs by less than one standard deviation (S1: $22.41$ vs $22.92$ MB/s, S5: $22.18$ vs $21.73$ MB/s, S8: $19.04$ vs $18.80$ MB/s).



\section{Confidence Intervals for Throughput and Lead Time}
\label{app:stats}

This appendix gives confidence intervals for the paper's two main claims, throughput gain and proactive lead time. Each of the ten scenarios (S1-S10) is run three times under the agent, reactive, and static policies. We treat each scenario's three-repeat mean as one independent observation, since run-to-run noise within a scenario is a separate concern from variation across scenarios. Intervals are Student $t$ intervals across scenarios, wide given the small sample. Table~\ref{tab:ci_summary} summarises both claims with their intervals and significance tests.

\begin{table}[t]
\centering
\caption{Confidence intervals and significance for the two main claims. The sampling unit is the scenario.}
\label{tab:ci_summary}
\footnotesize
\renewcommand{\arraystretch}{1.5}
\begin{tabularx}{0.8\columnwidth}{XccX}
\toprule
\textbf{Claim} & \textbf{Estimate} & \textbf{95\% CI} & \textbf{Test} \\
\midrule
Throughput gain, absolute & $4.01$\,MB/s & $[2.48, 5.54]$ & paired $t$, $p{=}0.0002$, $n{=}10$ \\
Throughput gain, relative & $40\%$ & $[28\%, 56\%]$ & scenario bootstrap \\
Proactive lead time & $34.2$\,s & $[26.8, 41.5]$ & $t$ interval, $n{=}8$ \\
\bottomrule
\end{tabularx}
\end{table}

\begin{table}[htbp]
\centering
\caption{Per-scenario mean throughput (MB/s, three repeats) and proactive lead time. Gain is agent minus reactive. Lead is the reactive minus agent reroute time.}
\label{tab:stats}
\begin{tabular}{lrrrrr}
\toprule
\textbf{Scenario} & \textbf{Agent} & \textbf{Reactive} & \textbf{Static} & \textbf{Gain} & \textbf{Lead (s)} \\
\midrule
S1  & 14.99 & 8.38  & 4.63  & $+6.61$ & 41.7 \\
S2  & 14.60 & 10.77 & 2.16  & $+3.82$ & 27.7 \\
S3  & 14.51 & 8.01  & 2.94  & $+6.50$ & 52.7 \\
S4  & 14.96 & 11.25 & 2.96  & $+3.71$ & 27.7 \\
S5  & 14.98 & 11.54 & 3.70  & $+3.45$ & 28.0 \\
S6  & 14.94 & 12.05 & 12.19 & $+2.88$ & --- \\
S7  & 6.83  & 6.89  & 4.70  & $-0.06$ & 31.7 \\
S8  & 13.74 & 11.29 & 4.97  & $+2.45$ & 31.4 \\
S9  & 14.69 & 7.98  & 6.36  & $+6.71$ & 92.3$^{\dagger}$ \\
S10 & 15.07 & 11.02 & 3.52  & $+4.05$ & 32.4 \\
\midrule
\textbf{Mean} & \textbf{13.93} & \textbf{9.92} & \textbf{4.81} & \textbf{+4.01} & \textbf{34.2} \\
\bottomrule
\end{tabular}
\end{table}

Table~\ref{tab:stats} lists the throughput and lead time for each of the ten scenarios individually. The $40\%$ relative gain reported earlier comes from the ratio of the two dataset-wide means. Additionally, averaging each scenario's own relative gain gives a consistent result, a mean of $42\%$ and a median of $34\%$. The small difference is because scenarios with higher throughput count for more in the dataset-wide ratio than in a plain average of ten percentages. Lead time is defined per scenario as the reactive reroute time minus the agent reroute time. Reactive never fires in S6, so no lead time exists there, and it fires in only two of three repeats in S9 (marked $\dagger$), so both scenarios are left out of the $34.2$\,s mean. S7 is the only scenario where the agent does worse than reactive, a $0.06$\,MB/s difference caused by the destination-selection failure discussed in Section~\ref{sec:results}.



\end{document}